\documentclass[12pt,preprint]{aastex}
\usepackage{graphicx,ulem}
\usepackage{color}
\usepackage{natbib}
\definecolor{green}{rgb}{0,.6,0}
\def\tr#1{#1}
\def\tra#1{#1}
\def\trb#1{#1}
\def\tb#1{}
\def\trc#1{#1}
\def\tbc#1{}

\begin{document}
\title{Magnetohydrodynamic Simulations of the Formation of Molecular Clouds toward the Stellar Cluster Westerlund 2: Interaction of a Jet with Clumpy Interstellar Medium}
\author{Yuta \textsc{Asahina}$^{1}$, Tomohisa \textsc{Kawashima}$^{1}$, Naoko \textsc{Furukawa}$^{2}$, Rei
\textsc{Enokiya}$^{2}$, Hiroaki \textsc{Yamamoto}$^{2}$, Yasuo \textsc{Fukui}$^{2}$ and Ryoji \textsc{Matsumoto}$^{3}$}

\email{asahina@cfca.jp}

\affil{$^{1}$National Astronomical Observatory of Japan, Osawa, Mitaka, Tokyo 181-8588, Japan}
\affil{$^{2}$Department of Physics, Nagoya University, Furo-cho, Chikusa-ku,
Nagoya 464-8602, Japan}
\affil{$^{3}$Department of Physics, Graduate School of Science, Chiba
University, 1-33 Yayoi-cho, Inage-ku, Chiba 263-8522, Japan}
\email{ }

%\maketitle

\begin{abstract}
The formation mechanism of CO clouds observed with NANTEN2 and Mopra telescope toward the 
stellar cluster Westerlund 2 is studied by three-dimensional magnetohydrodynamic (MHD) 
simulations taking into account the interstellar cooling.
These molecular clouds show a peculiar shape composing of an arc-shaped cloud in one side 
of a TeV $\gamma$-ray source HESS J1023-575 and a linear distribution of clouds (jet clouds) 
in another side. 
We propose that these clouds are formed by the interaction of a jet with interstellar neutral 
hydrogen (HI) clumps.
By studying the dependence of the shape of dense cold clouds formed by shock compression and 
cooling on the filling factor of HI clumps, we found that the density distribution of HI clumps determines the shape of molecular clouds formed by the jet-cloud interaction; arc-clouds are formed when the 
filling factor is large. 
On the other hand, when the filling factor is small, molecular clouds align with the jet. 
The jet propagates faster in models with small filling factors. 
\end{abstract}


\section{Introduction}
The CO J=1-0 observations 
toward the stellar cluster Westerlund 2 with NANTEN2 
telescope revealed peculiar shaped molecular clouds around a TeV $\gamma$-ray source HESS1023-575 \citep{2009PASJ...61L..23F}; 
a semi-circular molecular cloud (arc-like cloud) on one side of the TeV $\gamma$-ray source and a linear molecular cloud (jet-like cloud) on the other side. 
Subsequent high-resolution CO (J=2-1 and 1-0) observations with the NANTEN2 and Mopra telescopes provided significant details of their distribution at a factor of 2-5 higher resolution, confirming the arc- and jet-like clouds which are likely associated with HESS1023-575 \citep{2014ApJ...781...70F}. 
These authors argued that the clouds are 
further than Westerlund 2 by ~1 kpc 
so that the arc- and jet-like clouds and HESS1023-575 are not physically connected with Westerlund 2. 
An obvious explanation of the formation of the arc-cloud is a supernova explosion but the 
noncircular shape of the arc-cloud suggests that the explosion which formed the arc-cloud should be highly asymmetric. 
On the other hand, the 
correlation between the molecular clouds and the TeV $\gamma$-ray source indicates that the
microquasar jet ejected from the $\gamma$-ray source triggered the formation of the molecular clouds.
The formation of molecular clouds induced by the microquasar jet was discussed by \cite{2008PASJ...60..715Y},
in which they reported that molecular clouds found by NANTEN CO observations are aligned with
the radio and X-ray jet ejected from a microquasar SS433.

\tr{
\tr{H}ydrodynamic (HD) or magnetohydrodynamic (MHD) simulations 
have been conducted to study the interaction of astrophysical jets with the interstellar medium (ISM). 
\cite{1982A&A...113...285N} studied structures of a supersonic jet propagating into uniform intergalactic medium by two-dimensional (2D) axisymmetric HD simulations. 
They revealed that the propagating jets form structures such as a bow shock ahead of the jet, a jet terminal shock, internal shock, and cocoon. 
Axisymmetric 2D MHD simulations were conducted by \tr{a number of authors} \citep[e.g.,][]{1986ApJ...311L..63C, 1989ApJ...344...89L, 1990A&A...229..378K, 1992PASJ...44...245T}. 
\cite{1993ApJ...403...164} studied helical kink instability of a jet interacting a dense gas cloud by three-dimensional (3D) MHD simulations. 
These simulations were carried out by assuming adiabatic fluid. 
}

\tr{
Effects of the radiative cooling on the propagation of protostellar jets were studied by \cite{1989ApJ...337L..37B,1990ApJ...360..370B}. 
They carried out axisymmetric 2D simulations taking into account the radiative cooling in the temperature range $10^{4}\ \mathrm{K} < T < 10^{6}\ \mathrm{K}$. 
They showed that cooling enhanced by shock compression at the head of the jet \tr{forms} cool shell with temperature $T \sim 10^{4}\ \mathrm{K}$. 
Axisymmetric 2D MHD simulations of the protostellar jets 
taking into account the radiative cooling were carried out 
by \cite{1998ApJ...494L..79F}, \cite{2000ApJ...540..192S}, and \cite{2008A&A...488..429T}. 
}

\cite{2014ApJ...789...79A} presented the results of two-dimensional axisymmetric magnetohydrodynamic (MHD)
simulations of the interaction of a jet and neutral hydrogen (HI) cloud by taking
into account the interstellar cooling.
They adopted a cooling function \tr{applicable to the high density gas with $T<10^{4}\ \mathrm{K}$} \citep{2006ApJ...652.1331I}. 
\cite{2014ApJ...789...79A} showed that cooling instability 
triggered by the shock compression of HI \tr{layer} forms 
dense, cold ($<100\ \mathrm{K}$) \tr{arc-like} cloud 
along the bow shock ahead of the jet and \tr{elongated dense clouds} 
in the sheath surrounding the jet. 
The mass and the speed of the dense clouds are comparable to those
observed in clouds aligned with the SS433 jet.

It is possible that molecular clouds toward 
TeV $\gamma$-ray source HESS1023-575 
can be formed by the same mechanism. 
If both arc- and jet-clouds were formed by the same jet ejection event, the asymmetry of the molecular cloud distribution may be due to the difference in distributions of the HI clouds. 
In order to better understand the interaction of the jet and the ambient HI gas, we have carried out magnetohydrodynamic numerical simulations of the arc- and jet-clouds and present the results in this paper. 
We present our simulation models in Section 2 and numerical results in Section 3. 
Section 4 gives summary and discussion. 

\section{Numerical Model}
We conduct 3D MHD simulations in Cartesian coordinates $(x, y, z)$. 
The basic equations of ideal MHD are
\begin{equation}
\frac{\partial \rho}{\partial t}+{\bf \nabla}
 \cdot \left (\rho \mbox{\boldmath $v$} \right) =0
\end{equation}
\begin{equation}
\frac{\partial \left( \rho \mbox{\boldmath $v$} \right)}{\partial t}
 +{\bf \nabla} \cdot \left(\rho \mbox{\boldmath $v$} \otimes
 \mbox{\boldmath $v$} +p+\frac{B^{2}}{8\pi} -\frac{\mbox{\boldmath $B$} \otimes
 \mbox{\boldmath $B$}}{4\pi}    \right)=0
\end{equation}
\begin{equation}
\frac{\partial}{\partial t} \left( e+\frac{B^{2}}{8\pi} \right)+{\bf \nabla} \cdot
 \left[\left( e+p \right) \mbox{\boldmath $v$}
 -\frac{ \left(\mbox{\boldmath $v$} \times \mbox{\boldmath $B$}\right) \times
 \mbox{\boldmath $B$}}{4\pi} \right]=-\rho L
\end{equation}
\begin{equation}
\frac{\partial \mbox{\boldmath $B$}}{\partial t}={\bf \nabla}
 \times \left(\mbox{\boldmath $v$} \times \mbox{\boldmath $B$}\right)
\end{equation}
where $\rho, \mbox{\boldmath $v$},p,\mbox{\boldmath $B$}$, and $L$ are density, velocity,
pressure, magnetic field, and the cooling function, respectively, and
\begin{equation}
e=\frac{p}{\gamma -1}+\frac{\rho v^{2}}{2}
\end{equation}
is the energy density of the gas. We adopt the cooling function,
\begin{equation}
\rho L = n(-\Gamma +n\Lambda) \exp \left\{- \left[ \mathrm{max} \left(
  \frac{T}{7000}-1,0\right) \right] ^{4}\right\}
\end{equation}
\begin{equation}
\Gamma = 2 \times 10^{-26}\ \mathrm{ergs\ s^{-1}}
\end{equation}
\begin{eqnarray}
\Lambda & = & 7.3\times 10^{-21} \exp \left( \frac{-118400}{T+1500} \right) \nonumber \\
  & &+7.9\times 10^{-27}\exp \left( \frac{-200}{T} \right)\ \mathrm{ergs\ cm^{3}\ s^{-1}}
\end{eqnarray}
where $n$, $\Gamma$, and $\Lambda$ are number density, heating rate, and cooling rate, respectively.
Basic equations and cooling function for interstellar cooling are the same as those in \cite{2014ApJ...789...79A}.
\tr{In contrast to \cite{1989ApJ...337L..37B, 1990ApJ...360..370B} and \cite{1998ApJ...494L..79F} in which the radiative cooling in the temperature range $T>10^{4}\ \mathrm{K}$ is taken into account, we consider the cooling of the HI gas with temperature $T<10^{4}\ \mathrm{K}$. 
We cut off cooling in the temperature range $T>10^{4}\ \mathrm{K}$ 
because the jet internal plasma is continuously heated by internal shocks \citep{2014ApJ...789...79A}. 
Cooling can be significant in the sheath between the hot ($T>10^{6}$ K) jet and the cool ($T\sim 200$ K) HI cloud but the density and temperature distribution is essentially determined by the cooling in $T<10^{4}$ K. 
Inclusion of the cooling in $10^{4}$ K $<T< 10^{6}$ K only leads to the thinning of the interface between the hot jet and the cold, dense cloud formed by the cooling instability. 
In this paper, we neglect thermal conduction. 
Thermal conduction determines the width of the interface between the warm ($T \sim 10^{4}$ K) gas and the hot plasma ($T>10^{5}$ K). 
The thickness of the interface, called the Field's length \citep{1990ApJ...358..375B} is the order of 1 pc. 
In this paper, instead of resolving the interface in the temperature range ($10^{4}$ K $<T< 10^{6}$ K) taking into account thermal conduction, we approximate the interface by contact discontinuity where density and temperature jumps.
}

\tr{
\cite{2014ApJ...789...79A} studied the dependence of numerical results on the jet speed and showed that the speed of the working surface of the jet is proportional to the square root of the jet speed when the jet mass flux is fixed. 
The speed of the cold, dense clouds formed around the sheath between the jet and the ambient medium decreases with time, and approaches $0.5\ \mathrm{km\ s^{-1}}$. 
In \cite{2014ApJ...789...79A}, preexisting dense HI region is approximated by an uniform dense layer for simplicity. 
When the HI region has finite thickness, the speed of the working surface will increase. 
In this paper, we study the interaction of jets with spherical HI clouds. 
}

\begin{figure}[!ht]
   \plotone{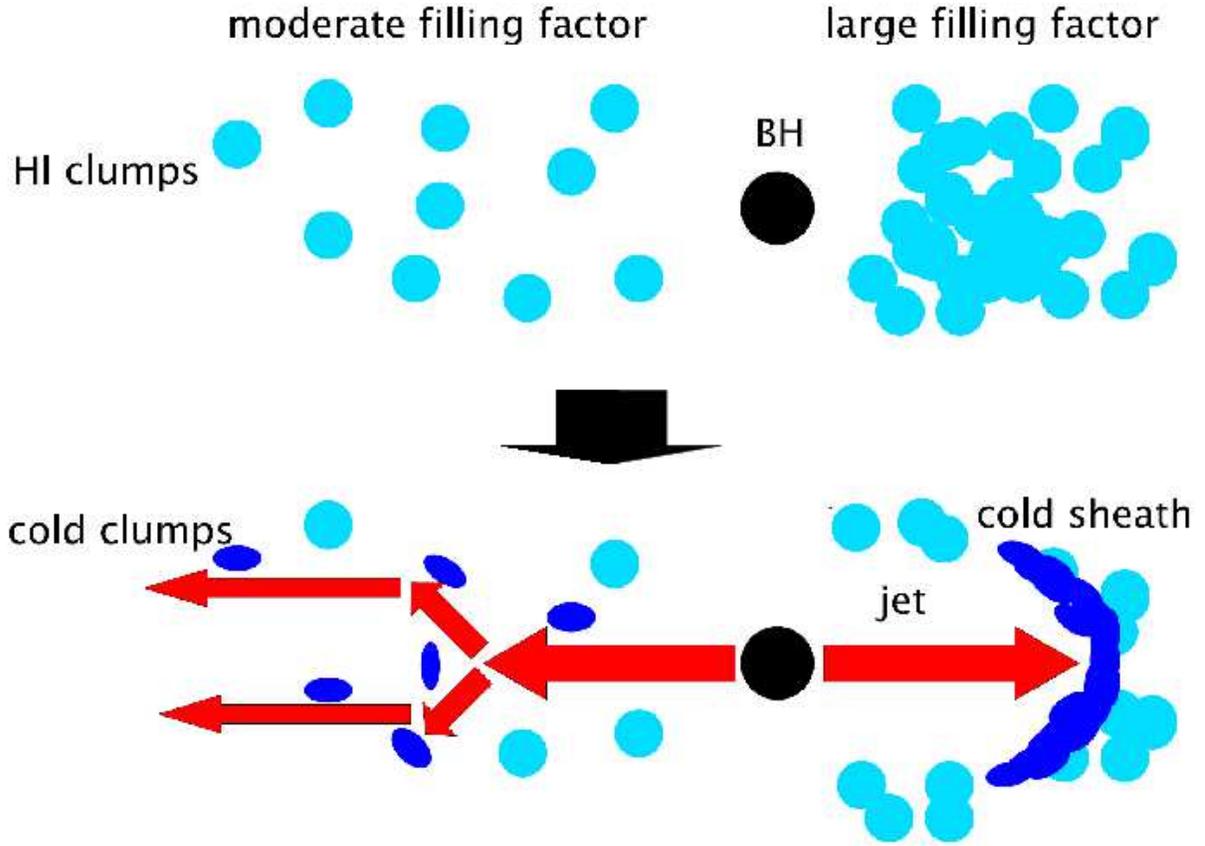}
  \caption{ Schematic picture of the interaction of a jet with the HI clumps. When the filling factor is small (left), the jet propagating in channels between the HI clumps forms jet-like molecular clouds. When the filling factor is large (right), the jet sweeps the HI clouds and forms an arc-shaped molecular cloud. \label{wl2f1}}
\end{figure}

\tr{
We carry out simulations for two models; A large HI cloud model and clumpy HI clouds model. 
The former model extends our previous 2D axisymmetric study on the interaction of a jet and dense HI layer \citep{2014ApJ...789...79A} by taking into account the finite size of the HI cloud and 3D effect\tr{s}.
\trc{We carry out simulations for a head on collision in which the center of the cloud locates on the jet axis, and an off-center collision. }
We expect that the arc-shaped cloud can be easily formed without assuming a supernova explosion in \trc{these} model\trc{s}. 
We confirm that the arc-shaped cloud can be formed in a finite-sized HI cloud via the 
same cooling instability mechanism with \cite{2014ApJ...789...79A}. 
The \tra{clumpy clouds} model is more realistic than the \trc{large HI cloud models}. 
We mainly carry out simulations by assuming this model to propose a unified model for the arc- and jet-cloud. 
Figure \ref{wl2f1} schematically shows the interaction of a jet with clumpy HI clouds. 
When the filling factor of the HI clouds is moderate, the jet will propagate along the channels between the HI clouds (left hand side of Figure \ref{wl2f1}). 
On the other hand, when the filling factor is large, an arc-like molecular cloud may be formed (right hand side of Figure \ref{wl2f1}).
}

\tr{For both models,} we inject a jet with a radius $ r_{\mathrm{jet}} = 1\ \mathrm{pc}$. 
The jet speed is $5.8\times 10^{2}\ \mathrm{km\ s^{-1}}$ \tr{(Mach 3)}. 
\tr{The temperature and number density of the jet} are $2 \times 10^{6}\ \mathrm{K}$ and $5\times 10^{-4}\ \mathrm{cm^{-3}}$, respectively. 
\tr{We determined the temperature and number density so that the jet is thermally stable \tra{even} \tr{when} the
cooling in $T>10^4$ K is considered.}
\tra{The mass flux of the jet, which is $1.4 \times 10^{18} \mathrm{g\ s^{-1}}$, is chosen such that it is comparable to the Eddington mass accretion rate for a stellar mass black hole (an order of magnitude smaller than that of the jet ejected from the microquasar SS433).} 
At the initial state, we assume pure toroidal magnetic field $B_{\phi} \propto \sin(r/r_{\mathrm{jet}})$ in the jet and \tr{$\beta = p_{\mathrm{gas}}/(B^{2}/8\pi)=100$ at $r=0.5\ \mathrm{pc}$}. 
\tr{The magnetic field is enhanced in the sheath in 2D simulations \citep{2014ApJ...789...79A}. 
To confirm that the magnetic field \tr{is amplified} in 3D simulation \tr{as well as in 2D simulations}, 
we assume the same condition for the initial magnetic field. }
The number density and temperature of the HI cloud are $n_{\mathrm{HI}} = 6.9\ \mathrm{cm^{-3}}$ and, $T_{\mathrm{HI}}=200\ \mathrm{K}$, respectively. 
The warm ISM with the number density of $0.15 \ \mathrm{cm^{-3}}$ and the temperature of $9.3 \times 10^{3}\ \mathrm{K}$ is \tra{assumed to be} in pressure equilibrium with the HI cloud. 

\begin{figure}[!ht]
  \plotone{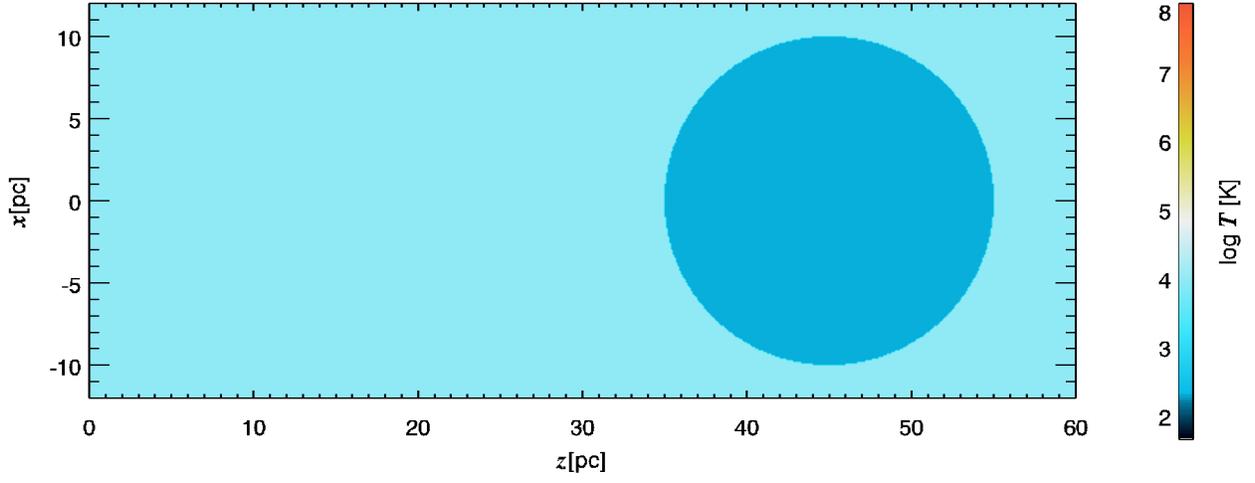}
  \caption{Temperature distribution in the initial state in $y=0$ plane.\
 We put the large spherical HI cloud. \label{wl2f2}}

\end{figure}

For a large cloud model, we assume a spherical HI cloud \trc{whose center is} at \trb{$(x,z)=(0\ \mathrm{pc}, 45\ \mathrm{pc})$} \trb{for \trc{a} head-on collision model.
  \trc{For the off-center collision model, the center of the cloud locates at} $\trb{(x,z)=(2.5\ \mathrm{pc}, 45\ \mathrm{pc})}$ }.
\trc{The radius of the cloud is $10\ \mathrm{pc}$.}
Figure \ref{wl2f2} shows the temperature distribution in the initial state in $y=0$ plane \trb{for the head-on collision model}. 
The size of the simulation region is $(L_{x}, L_{y}, L_{z}) = (24\ \mathrm{pc}, 24\ \mathrm{pc}, 60\ \mathrm{pc})$ and the number of grid points is $(N_{x}, N_{y}, N_{z}) = (240, 240, 600)$. 
For clumpy HI clouds model, we place spherical HI clumps with radii of 2 pc randomly in the region $z>10\ \mathrm{pc}$. 
\tr{When an HI clump overlaps with other HI clump, we set the number density as $n=6.9\ \mathrm{cm^{-3}}$ (i.e. density is not simply added) in such region in order to satisfy the condition for the thermal equilibrium. }
The volume filling factor of the HI clumps, defined as the ratio $V_{\mathrm{HI}}/V_{\mathrm{total}}$, is assumed to be 0.2, 0.8, and 0.9, \tr{for} models F02, F08, and F09, respectively. 
\tr{For the clumpy cloud model with filling factor 0.2, we also carried out simulations for Mach 4 jet (model F02H) and Mach 2 jet (model F02L). 
The jet velocities are $7.7 \times 10^{2}\ \mathrm{km\ s^{-1}}$ for model F02H and $3.9 \times 10^{2}\ \mathrm{km\ s^{-1}}$ for model F02L. 
At the initial state for models F02L and F02H, we assume the same density distribution as that for model F02. }
The size of the simulation region \tra{for the clumpy clouds model} is $(L_{x}, L_{y}, L_{z}) = (40\ \mathrm{pc}, 40\ \mathrm{pc}, 60\ \mathrm{pc})$ and the number of grid points is $(N_{x}, N_{y}, N_{z}) = (400, 400, 600)$. 
For both models, 
the grid size is $(\Delta x, \Delta y, \Delta z) = (0.1\ \mathrm{pc}, 0.1\ \mathrm{pc}, 0.1\ \mathrm{pc})$. 
We impose a symmetric boundary condition at $z=0$, and the other boundaries are assumed to be free boundaries. 
We apply an MHD code based on the HLLD Riemann solver \citep{2005JComp...208...315} with fifth-order spatial accuracy. 
The spatial accuracy is achieved by applying the WENO-Z \citep{2008JCoPh.227.3191B} and Monotonicity-Preserving (MP) schemes \citep{1997JCoPh.136...83S} adopted in \cite{2015ApJ...808...54M}.

\section{Numerical Results}
\subsection{Results for the Large Cloud Model}
\begin{figure}[!ht]
  \plotone{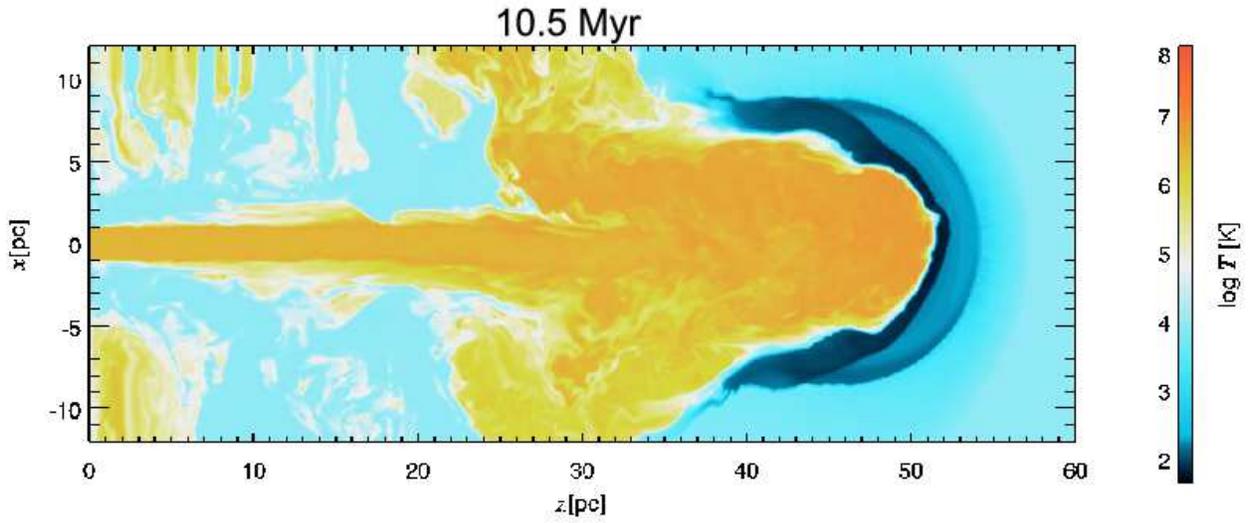}
  \caption{Temperature distribution at $t=10.5\ \mathrm{Myr}$ in $y=0$ plane.\
 The arc-like cold, dense sheath is formed.  \label{wl2f3}}
\end{figure}

Figure \ref{wl2f3} shows a result for the \trb{head-on collision} model at $t=10.5\ \mathrm{Myr}$ in $y=0$ plane. 
The shock-compressed HI cloud is cooled to about 50 K by cooling instability, and a cold, dense sheath, indicated by dark blue \tr{region}, is formed. 
\tr{This result is similar to that of 2D simulations in which \tr{infinitely thick} HI layer is assumed.} 
\tr{The cold, dense sheath is the thinnest around the head of the jet where thickness 
is $0.6-0.7\ \mathrm{pc}$. 
The sheath is resolved by 6-7 grids. 
The width of the interface between the cool sheath 
and warm gas where $10^{3}\ \mathrm{K}<T<10^{4}\ \mathrm{K}$ is determined by Field's length of the warm gas ($\sim 0.1\ \mathrm{pc}$). 
It is out of the scope of this paper to carry out simulations including thermal conductivity to resolve the Field's length. 
In this simulation, although we do not consider the heat conduction, the interface width is comparable to $0.1\ \mathrm{pc}$ 
because the grid size is $0.1\ \mathrm{pc}$. 
The interface width can become thinner 
if we carry out higher resolution simulations,
but the density of the cold sheath will not change 
because 
the thickness of the cold sheath is \tr{sufficiently} larger than the Field's length of the warm gas. 
As \cite{2014ApJ...789...79A} showed, the magnetic field is enhanced in the sheath \tr{to the} strength 6-7 times larger than that injected into the boundary at $z=0$,
which corresponds to $\beta \sim 20$.
The effect of the magnetic field is small since the plasma beta is higher than 1.}
The region between the jet terminal shock and the jet-cloud interface becomes turbulent, and the cocoon expands in the direction perpendicular to the jet axis. 
This radial expansion produces an arc-like cold, dense cloud. 

\begin{figure}[!ht]
  \plotone{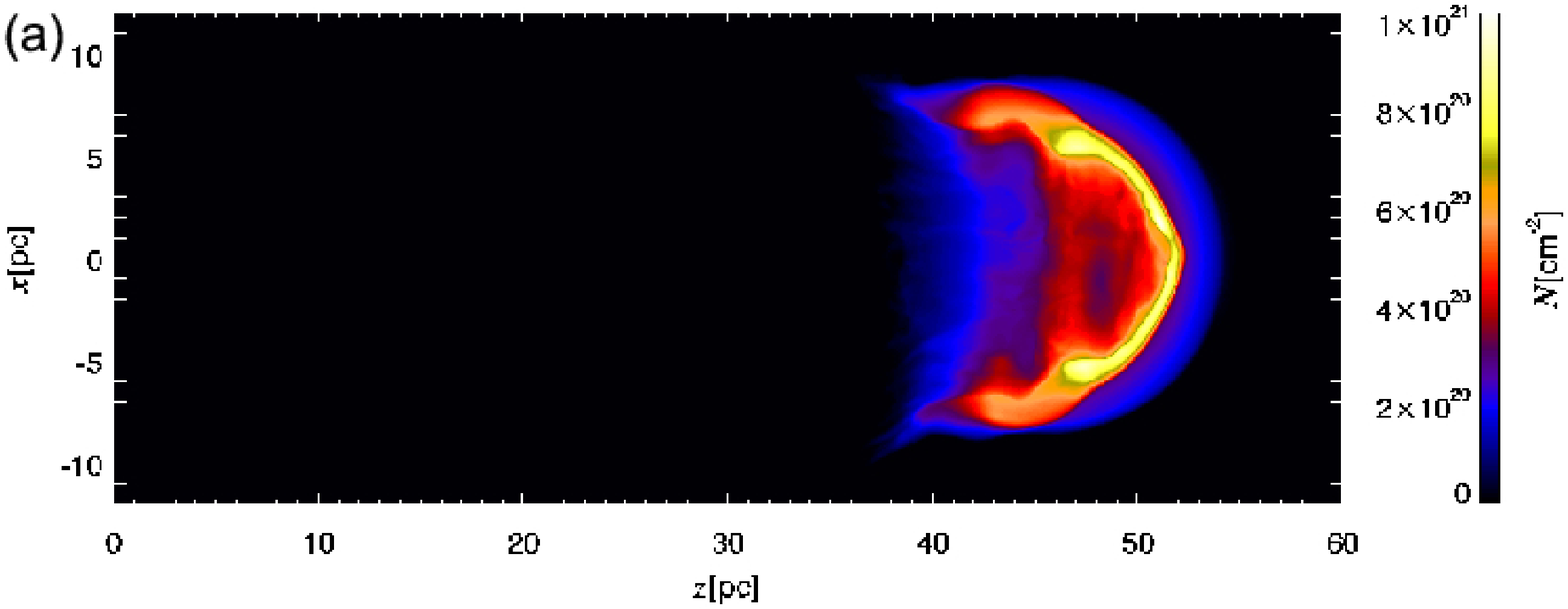}
  \plotone{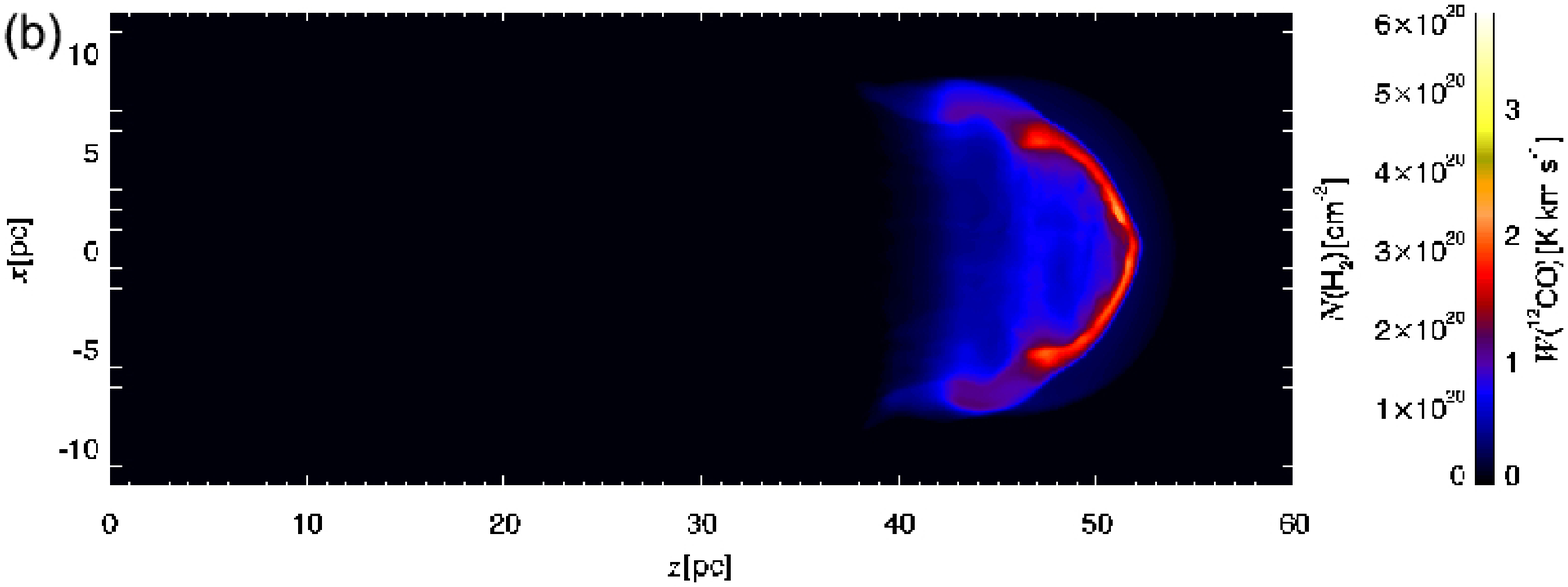}
  \caption{(a) Column number density of atomic hydrogen and (b) column number density of $\mathrm{H_{2}}$ and the CO intensity at $t=10.5\ \mathrm{Myr}$ observed from the $+x$ direction. For both column number densities, the peak appears ahead of the jet. The peak of the column number density is the order of $10^{21}\ \mathrm{cm^{-2}}$ and the peak of the $\mathrm{H_{2}}$ column number density is about $0.4 \times 10^{21}\ \mathrm{cm^{-2}}$  \label{wl2f6}}
\end{figure}

We computed the column number density of atomic hydrogen [Figure \ref{wl2f6} (a)] and the $\mathrm{H_{2}}$ column number density [Figure \ref{wl2f6} (b)] observed from the $+x$ direction. 
\tr{To compute the $\mathrm{H_{2}}$ number density,
we assume solar abundance and neglect background UV radiation \citep[e.g.,][]{2014MNRAS.440.3349R} 
and estimate 
$2n_{\mathrm{H_{2}}}/n$ in $10\ \mathrm{K} < T < 200\ \mathrm{K}$ 
from Figure 4 in \cite{2014MNRAS.440.3349R} 
\tr{assuming $\log (2n_{\mathrm{H_{2}}}/n) \propto \log n$ and interpolating for $n$ between $2n_{\mathrm{H_{2}}}/n =1$ when $n>10^{2}\ \mathrm{cm^{-3}}$ and $2n_{\mathrm{H_{2}}}/n =0.5 \mathrm{min}\left[1,\left(T/50\right)^{-1.16}\right]$ when $n=1\ \mathrm{cm^{-3}}$. }
This method is the same as in \cite{2014ApJ...789...79A}. 
}

We used the X-factor $N(\mathrm{H_{2}})/W(^{12}\mathrm{CO}(J=1-0))=1.6\times 10^{20}\ \mathrm{cm^{-2}/(K\ km\ s^{-1})}$ \citep{1997ApJ...481..205H} adopted in \cite{2014ApJ...781...70F}. 
The column number density is high in the sheath. 
The peak column number densities \tr{for the atomic hydrogen} and molecular hydrogen ahead of the jet are $N \sim 1 \times 10^{21}\ \mathrm{cm^{-2}}$ and $N(\mathrm{H_{2}}) \sim 0.4 \times 10^{21}\ \mathrm{cm^{-2}}$, \tr{respectively}. 
The peak of the integrated intensity of the HI $21\ \mathrm{cm}$ line is about $10^{3}\ \mathrm{K\ km\ s^{-1}}$ \citep{2005ApJS..158..178M}. 
The column number density \tr{for the atomic hydrogen} is estimated to be $1.8 \times 10^{21}\ \mathrm{cm^{-2}}$ by using the conventional factor $1.8 \times 10^{18} \mathrm{cm^{-2}/(K\ km\ s^{-1})}$.
The peak column number density in the simulations is comparable to that in HI observations. 
The peak $\mathrm{H_{2}}$ column density is smaller than the value of $N(\mathrm{H_{2}}) \sim 2.7 \times 10^{21}\ \mathrm{cm^{-2}}$ obtained by CO observations. 
This is because the width of the high column density region in our simulations ($\sim 10\ \mathrm{pc}$) is about half that of the observation. 
\tr{Furthermore,} the velocity dispersion is about $2\ \mathrm{km\ s^{-1}}$ in the simulation, which is about half that in the observation. 

\begin{figure}[!ht]
  \plotone{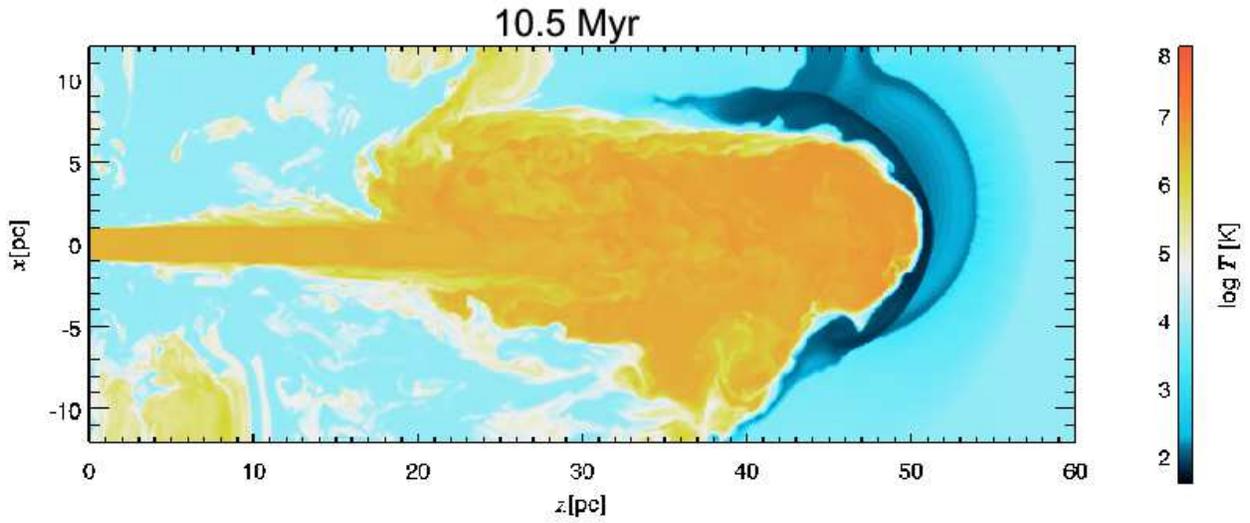}
  \caption{\trb{Temperature distribution at $t=10.5\ \mathrm{Myr}$ in $y=0$ plane for the off-center collision model. The arc-like cold, dense sheath is formed although the initial \trc{density} distribution is not axisymmetric.}  \label{lc2lte}}
\end{figure}

\begin{figure}[!ht]
  \plotone{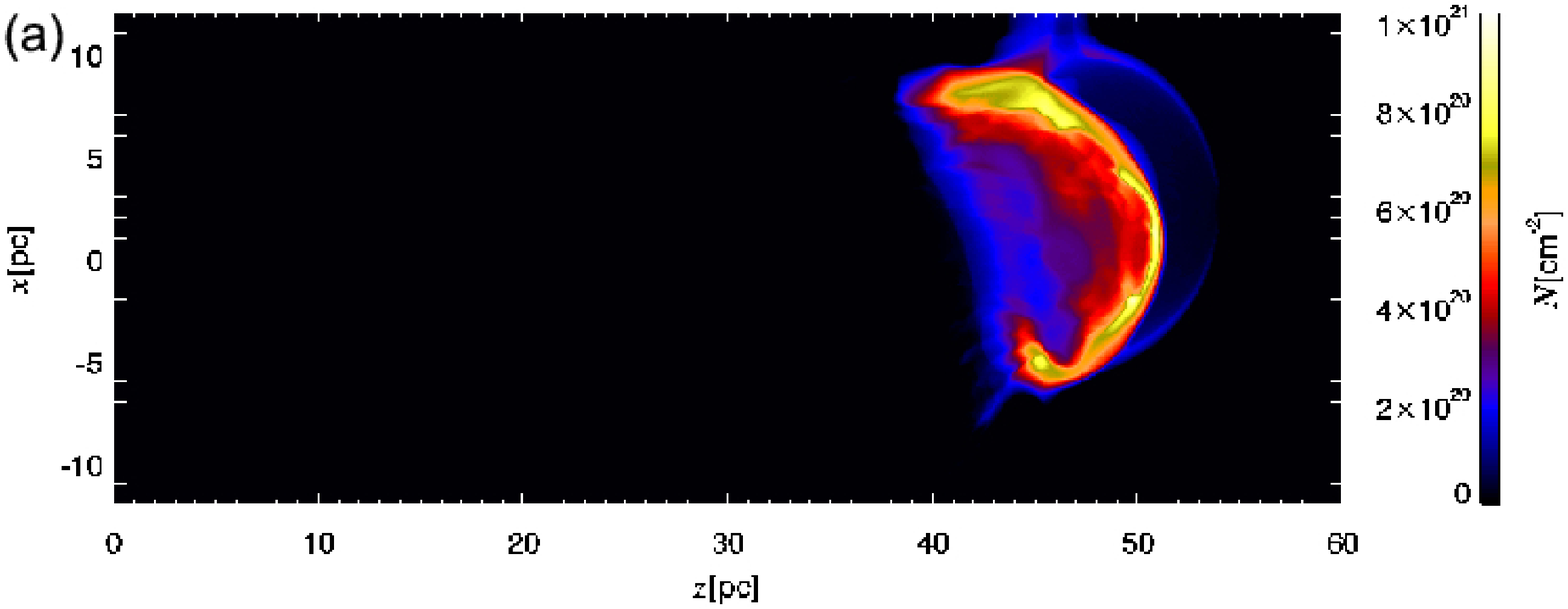}
  \plotone{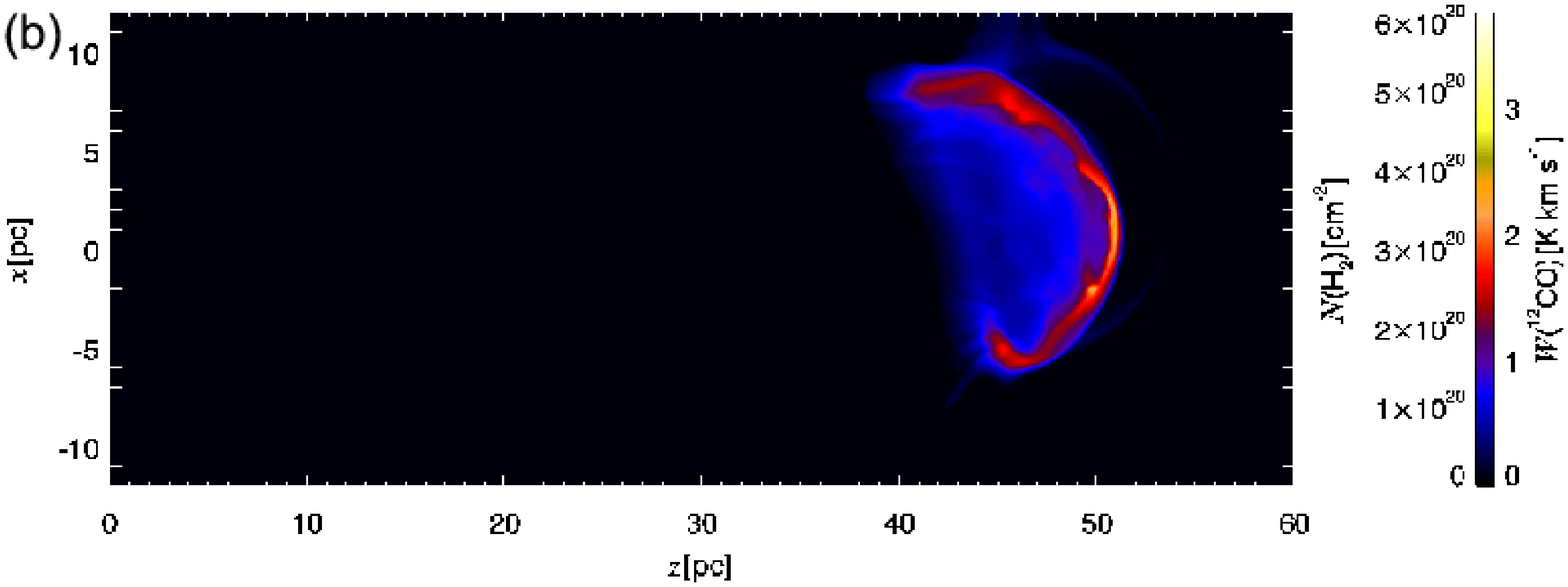}
  \caption{\trb{(a) Column number density of atomic hydrogen and (b) column number density of $\mathrm{H_{2}}$ and the CO intensity at $t=10.5\ \mathrm{Myr}$ observed from the $-y$ direction for the off-center collision model. For both column number densities, the peak \trc{of the column number density} appears ahead of the jet. The peak \trc{value} of the column number density is almost the same as \trc{those in} the head-on collision model.}  \label{lc2sig}}
\end{figure}

\trb{
  Figure \ref{lc2lte} shows the temperature distribution for the off-center collision model at $10.5\ \mathrm{Myr}$ in $y=0$ plane.
  The HI cloud cools down \trc{as the cloud is compressed by the jet.}
  \trc{Arc-like cold, dense region is formed in the off-center collision model as well as in the head-on collision model. 
    Figure \ref{lc2sig} shows the distribution of the column density for the off-center collision model.
    Although some asymmetry appears, the peak column densities are almost the same as those for the head-on collision model.
    These numerical results indicate that the arc-like cloud can be formed even when the jet collides with a large cloud at off-axis.
  }
}

\subsection{Results for Clumpy HI Clouds Model}
\begin{figure}[!ht]
  \epsscale{0.7}
  \plotone{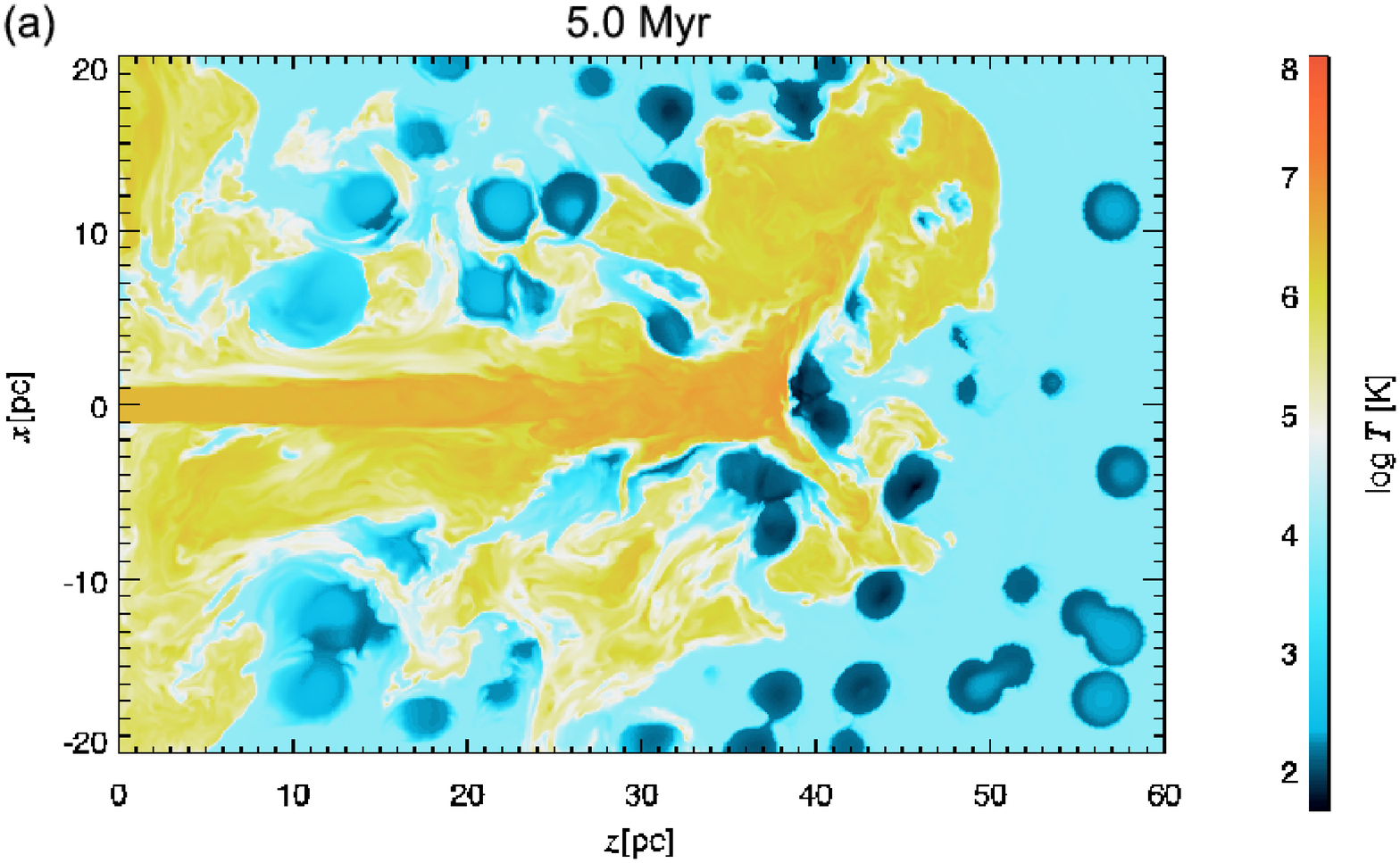}
  \plotone{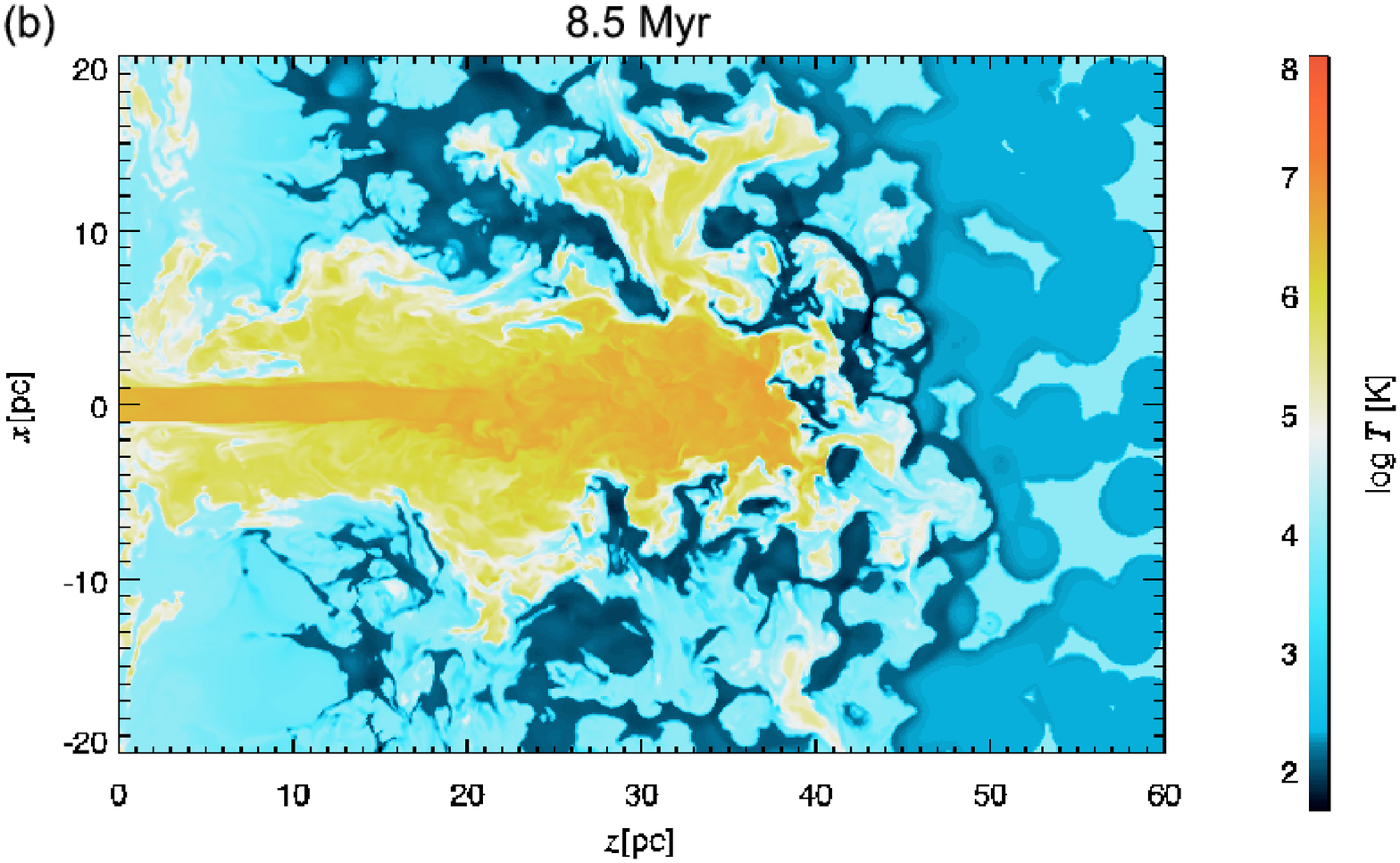}
  \plotone{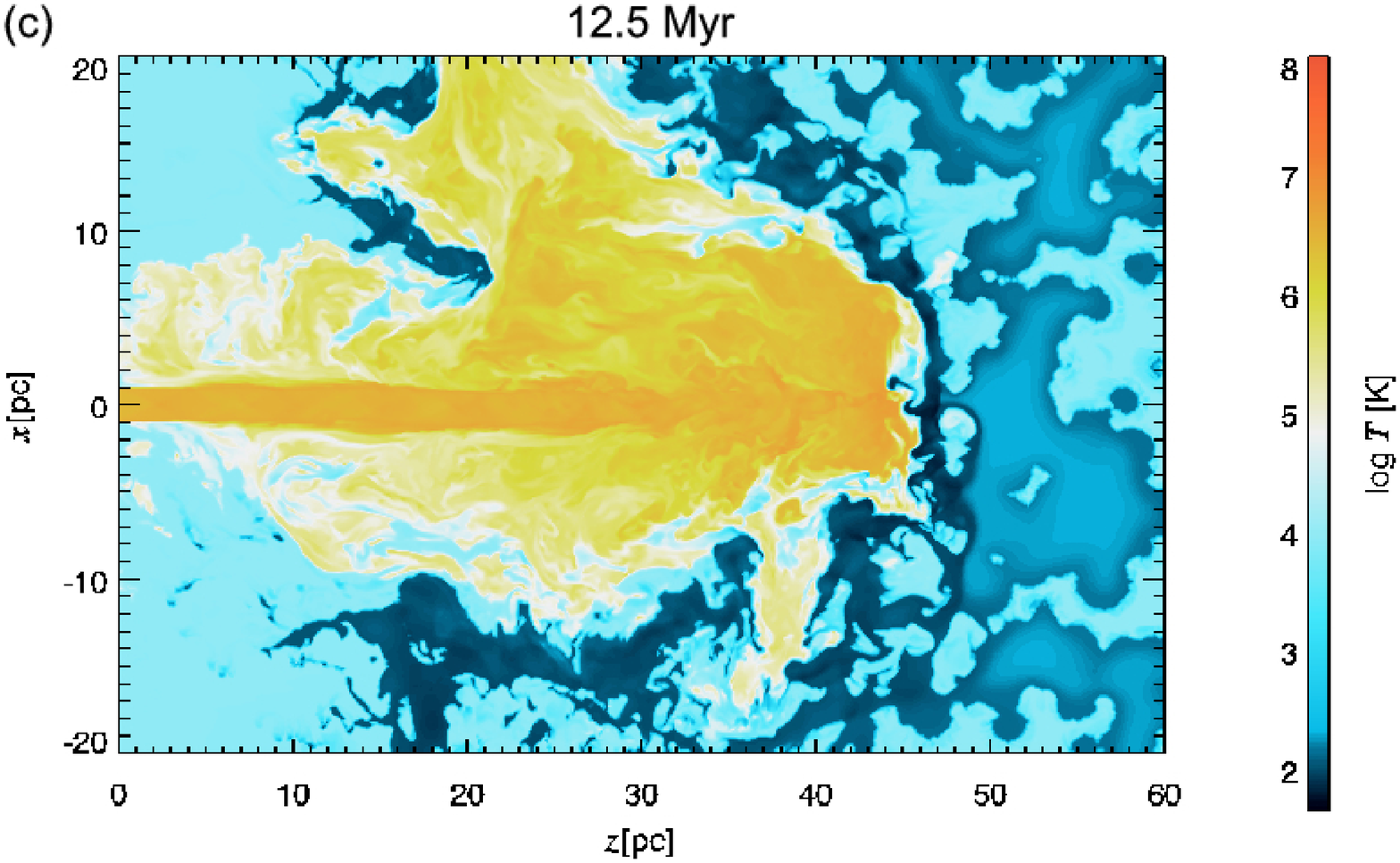}
  \caption{Temperature distribution for (a) model F02, (b) model F08, and (c) model F09. Dark blue shows the cold, dense clumps. The jet breaks up into branches especially for model F02.  \label{wl2f8}}
\end{figure}

Figure \ref{wl2f8} (a), (b), and (c) show the results for model F02 at $5.0\ \mathrm{Myr}$, model F08 at $8.5\ \mathrm{Myr}$, and model F09 at $12.5\ \mathrm{Myr}$, respectively. 
Since the jet propagates along the channels between the HI clumps, it breaks up into branches, and cold, dense clumps, indicated by dark blue, are formed by shock compression. 

The maximum number density is $150-300\ \mathrm{cm^{-3}}$ for all the models. 
The number density is high in the region where HI clumps are compressed by the jet. 
As time goes on, the high density region moves in the $+z$ direction. 
At the sides of the jet, the number density of the condensed gas decreases to $20-30\ \mathrm{cm^{-3}}$. 
This result is consistent with the observation that the CO emission is weaker near the TeV $\gamma$-ray source. 
For all the models, the maximum number density gradually decreases with time because the turbulence in the head of the jet decreases the ram pressure at the working surface. 

\begin{figure}[!th]
\epsscale{0.7}
    \plotone{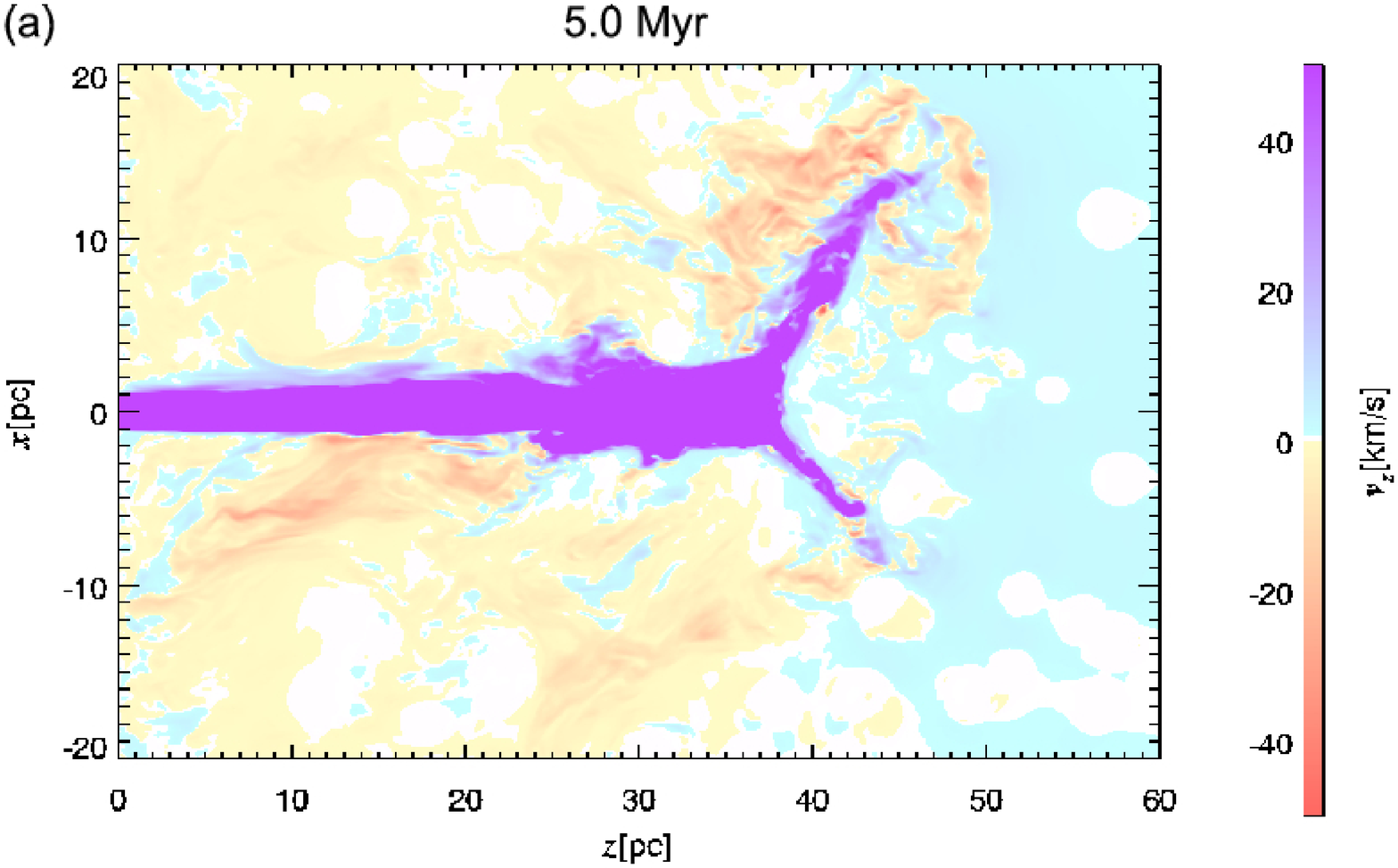}
    \plotone{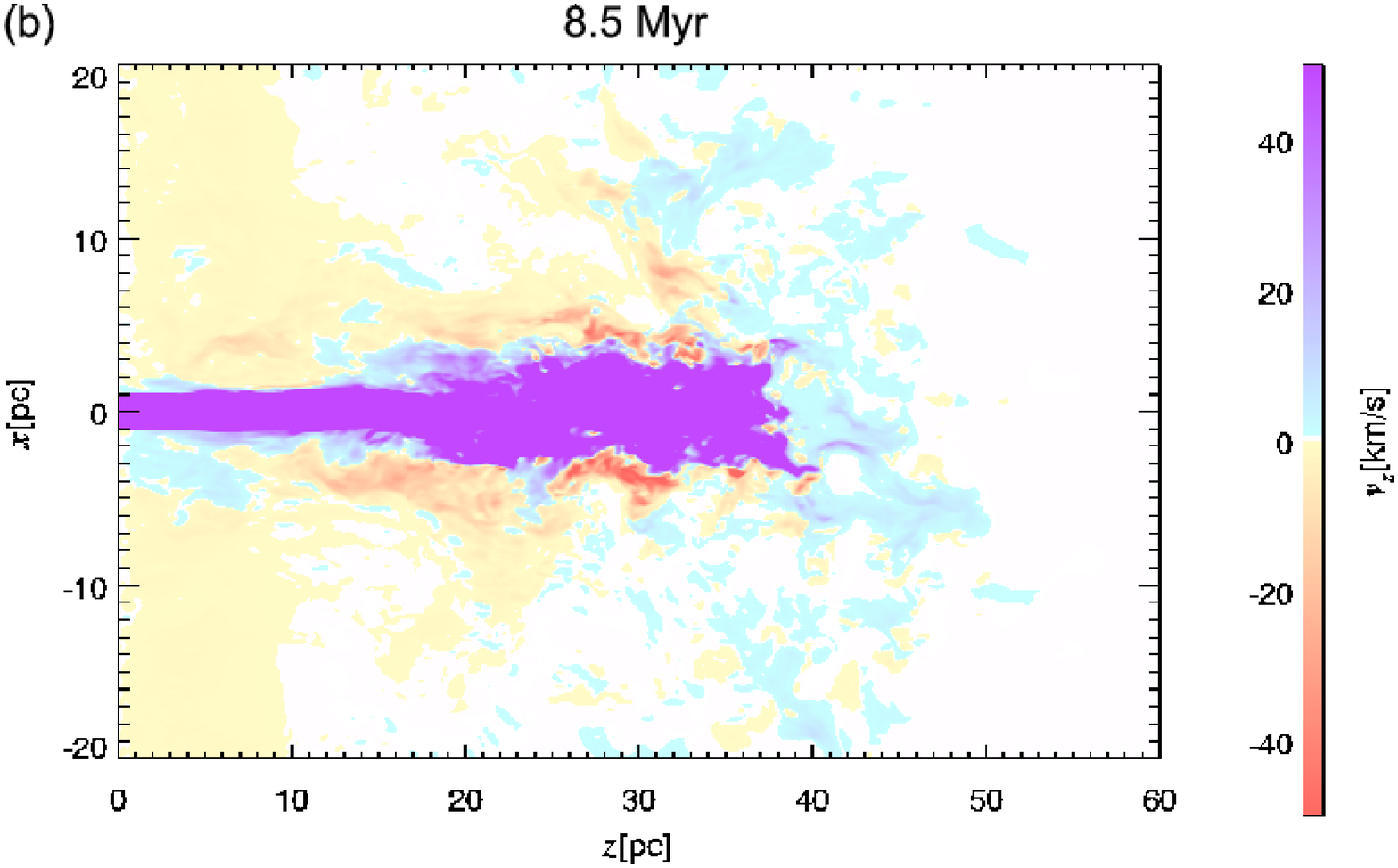}
    \plotone{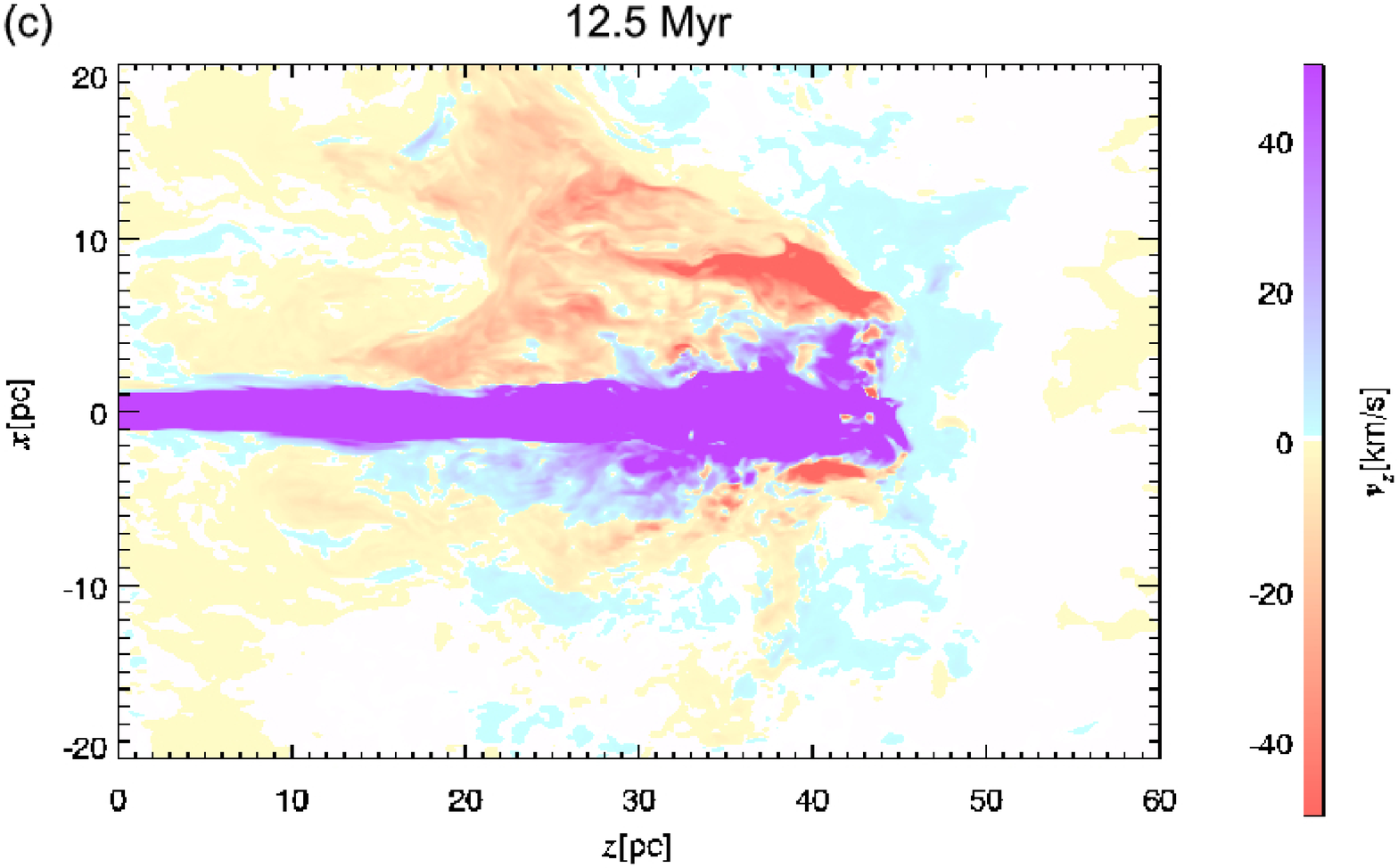}
    \caption{Distribution of the axial velocity $v_{z}$ for (a) model F02, (b) 
model F08, and (c) model F09 in $y=0$ plane. Violet shows the beam and red show
s the backflow. }
    \label{wl2f9}
\end{figure}

Figure \ref{wl2f9} shows the $v_{z}$ distribution in $y=0$ plane. 
\tr{The jet changes its direction after the jet collides with a clump at $z=35\ \mathrm{pc}$ for model F02 and F08, and $z=45\ \mathrm{pc}$ for model F09.}

In the early stage, the interface between the jet ($v > c_{\mathrm{s,jet}} \sim 200\ \mathrm{km\ s^{-1}}$) and the ambient gas is sharp for the velocity \tr{distribution}, where $c_{\mathrm{s,jet}}$ is the sound speed of the jet in the injection region. 
In the later stage, the interface is broadened by turbulence excited around the head of the jet. 
The beam having a velocity comparable to the injection velocity of the jet extends to $z=30\ \mathrm{pc}$, $35\ \mathrm{pc}$, and $40\ \mathrm{pc}$ for models F02, F08, and F09, respectively, in the last stage.  
These positions correspond to the position at which the jet collides with HI clumps (see Figure \ref{wl2f8}). 
Since the jet becomes turbulent after the collision with a clump, the jet speed gradually decreases and forms a broader interface for the velocity between the jet and the ambient medium.  

\begin{figure}[!th]
    \plotone{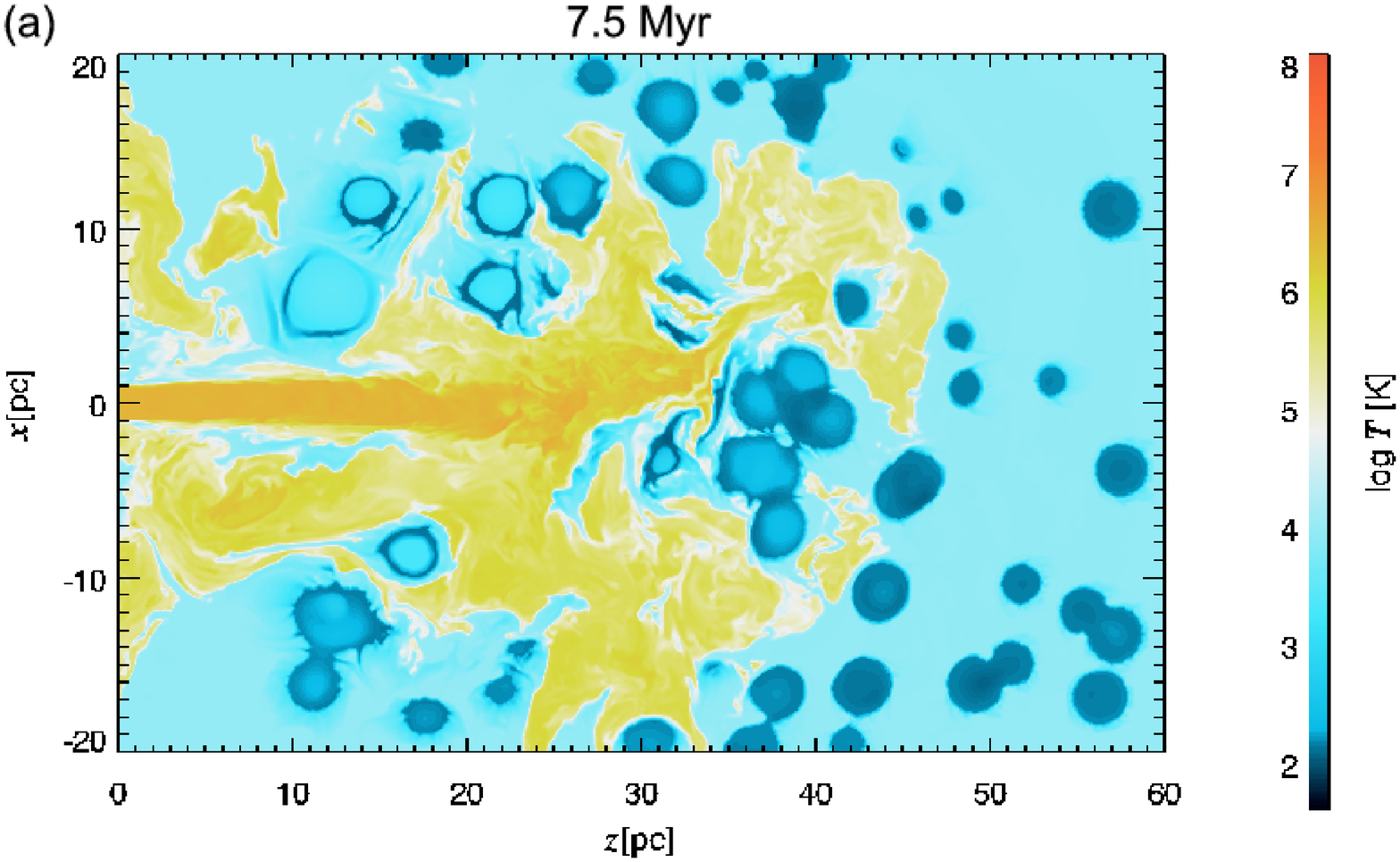}
    \plotone{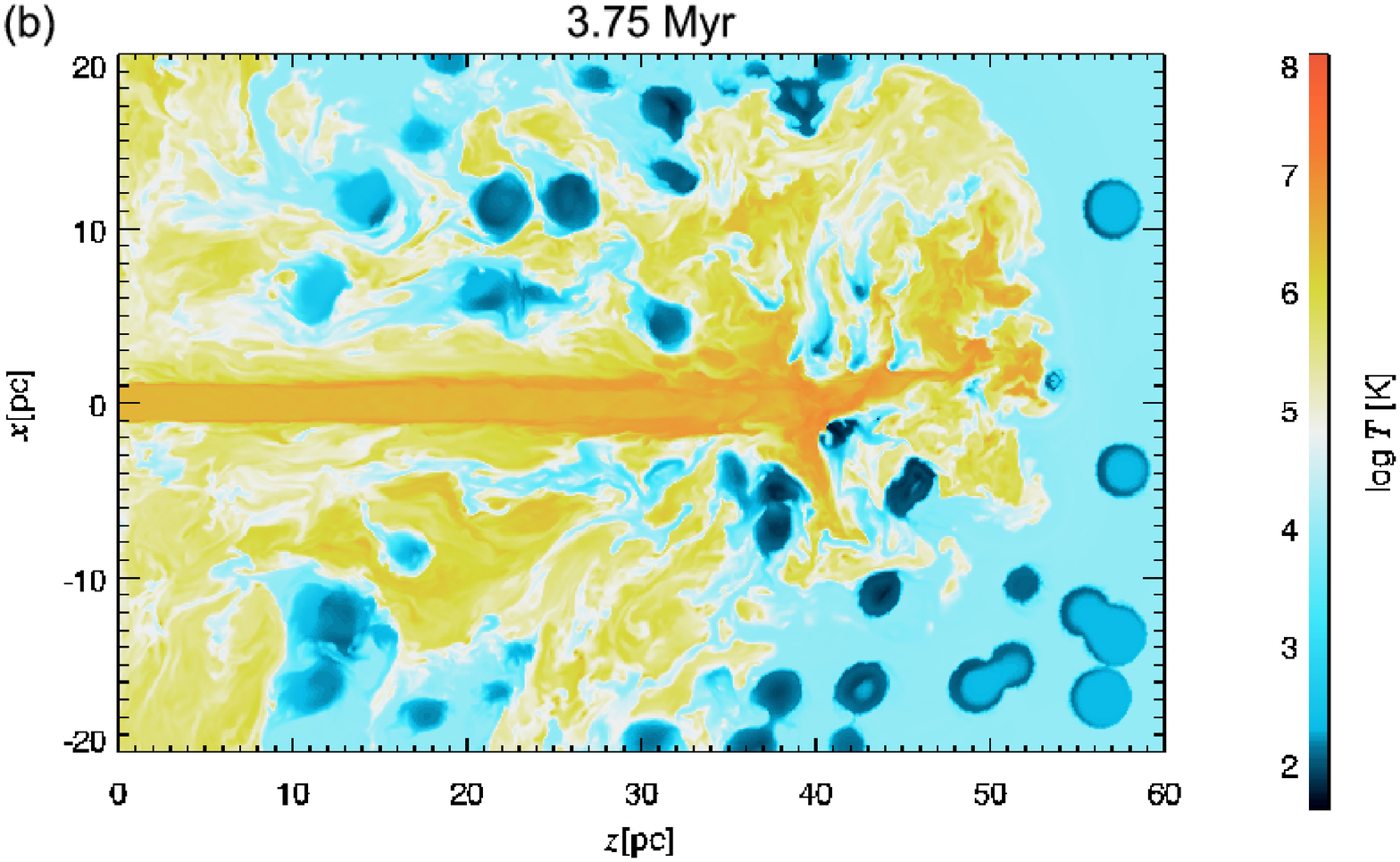}
    \caption{\tr{Temperature distribution of the result for (a) model F02L and (b) model F02H. For model F02L, the beam breaks up at $z=25$ pc. For model F02H, it is hard for the jet to be deflected by collision with the HI clumps.}}
    \label{f02hlte}
\end{figure}

\tr{
Figure \ref{f02hlte} (a) shows the temperature distribution for model F02L at $t=7.5\ \mathrm{Myr}$. 
The beam breaks up at $z=25$ pc and the jet is deflected at $z=35$ pc. 
On the other hand, the beam extends to $z=40$ pc for model F02H [see Figure \ref{f02hlte} (b)]. 
The deflection of the jet by the clump is smaller in model F02H than model F02. 
Since the jet power becomes large \tr{for jets with higher speed}, the bow shock \tr{becomes stronger} and the ram pressure \tr{increases}. 
The colder, denser clumps are formed in model F02H. 
For example, the maximum number densities are about $500\ \mathrm{cm^{-3}}$ for model F02H at $t=3.75$ Myr, $250\ \mathrm{cm^{-3}}$ for model F02 at $t=5.0$ Myr, and $30\ \mathrm{cm^{-3}}$ for model F02L at $t=7.5$ Myr. 
The minimum temperatures are about $30$ K for model F02H, $50$ K for model F02, and $100$ K for model F02L.
}

\begin{figure}[!th]
    \plotone{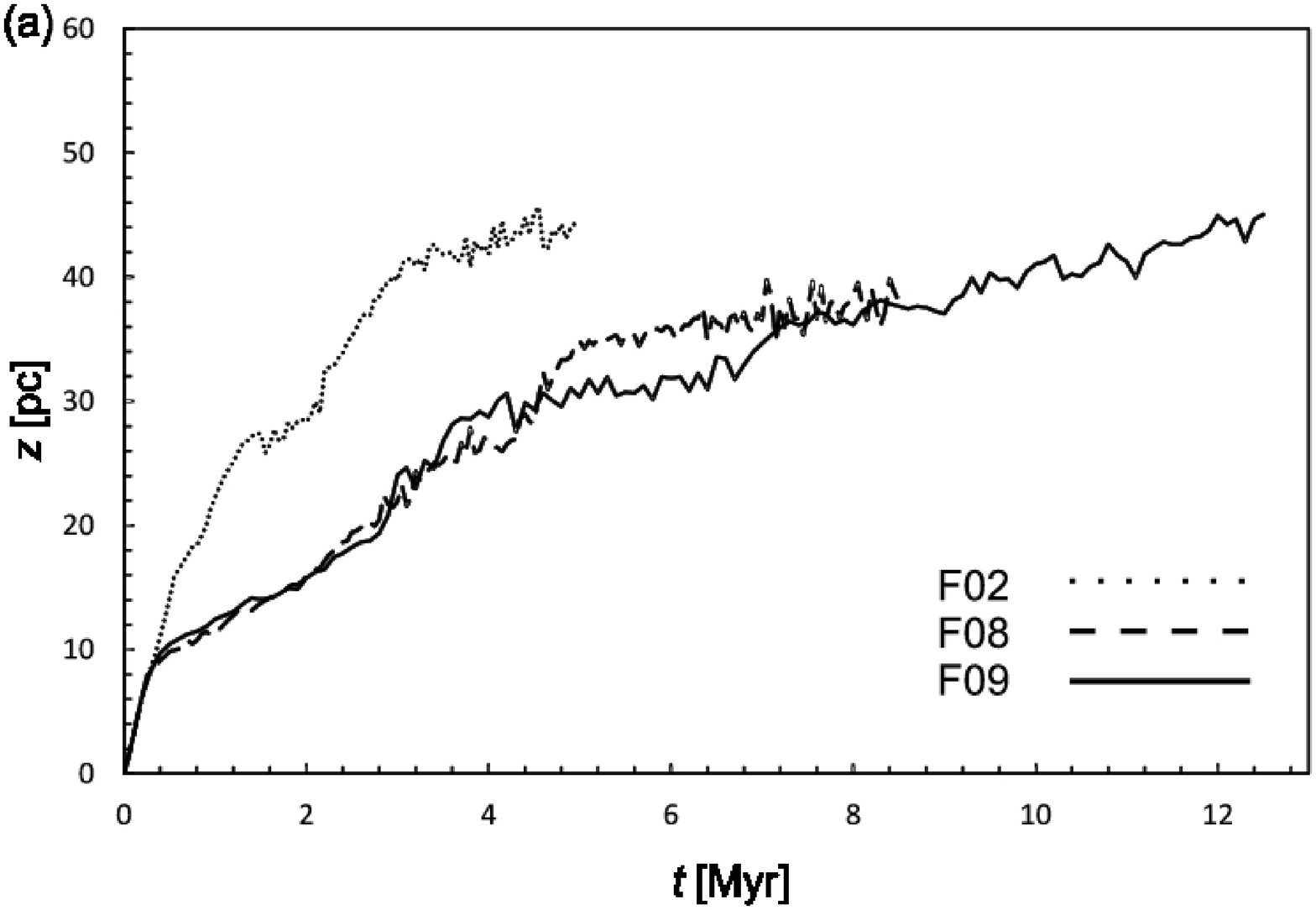}
    \plotone{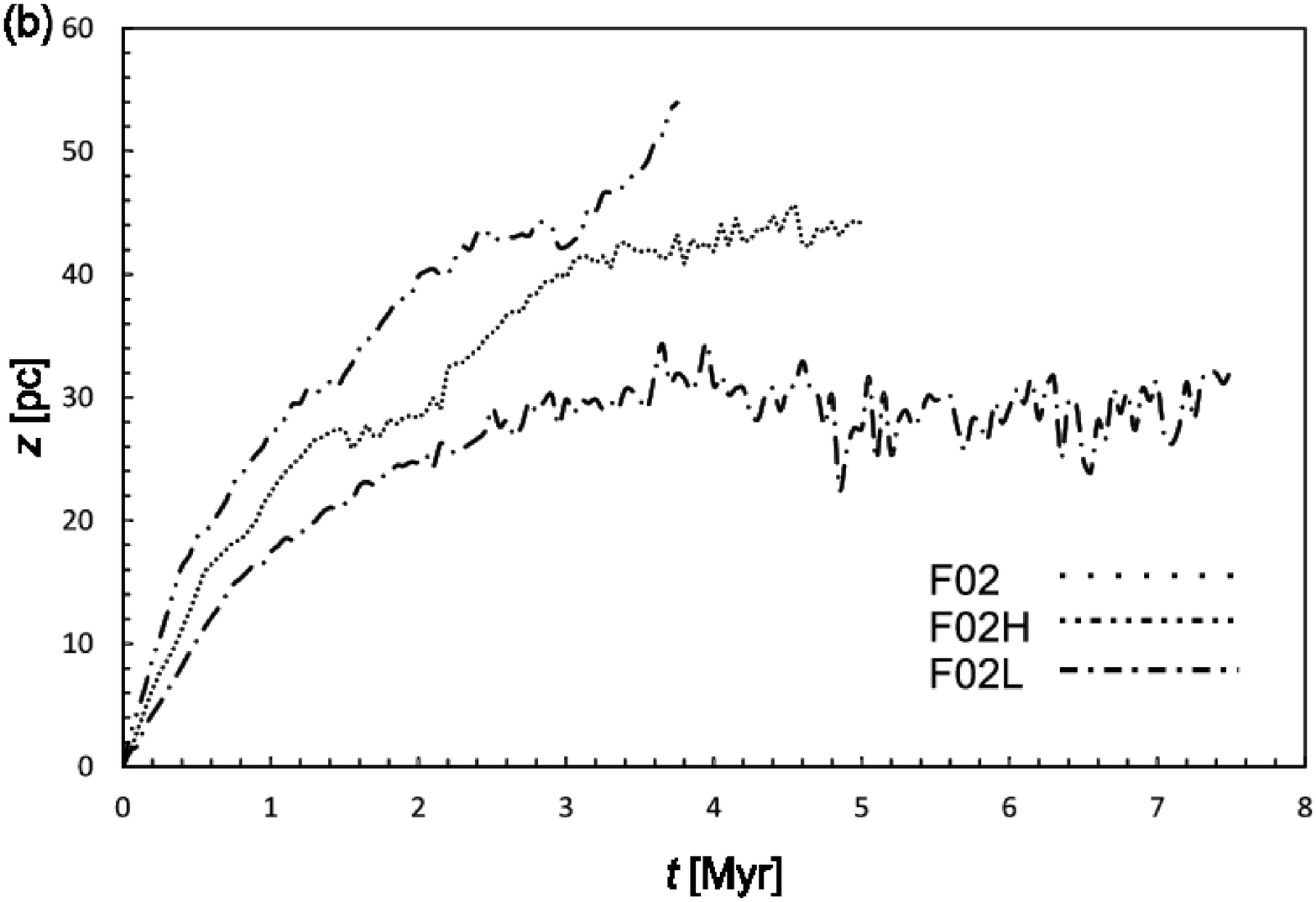}
    \caption{Propagation of the jet \tr{for (a) models with \tra{the} same jet speed and (b) models with different jet speed}. The vertical axis shows the position of the front end of the jet where the velocity exceeds $200\ \mathrm{km\ s^{-1}}$. The jet propagates through the warm ISM before the jet collides with the HI clumps. The jet for model F02 propagates faster than that for model F08 and F09.}
    \label{t-zp}
\end{figure}

Figure \ref{t-zp} \tr{(a) and (b)} plot the maximum $z$ \tr{of} the region where \tr{$v>200\ \mathrm{km\ s^{-1}}$}.
The curves coincide \tr{for models F02, F08 and F09} when $z<10\ \mathrm{pc}$ because the jet propagates in the region without HI clumps.
For model F02, the jet propagates faster than \tr{that for} models \tr{F08 and F09} since the filling factor is small\tr{er for model F02}.
However, the jet speed for model F08 is not always faster than that for model F09 because the HI clumps are located randomly in the initial state. 
\tr{For model F02L, the propagation of the beam stops since the beam breaks at $z=30$ pc. }

The propagation speed of the jet depends on the density distribution of the ISM. 
Since the dynamical pressure of the jet is equal \tr{to} that of the ambient medium 
with density $\rho_{\mathrm{ISM}}$ and that of the HI clump with density $\rho_{\mathrm{HI}}$ in a rest frame of the working surface \citep[e.g.,][]{1992PASJ...44...245T}, 
we estimate the propagation speed of the working surface as 
\begin{equation}
v_{\mathrm{ws, HI}} = v_{\mathrm{jet}}\frac{r_{\mathrm{jet}}}{r_{\mathrm{ws}}}\sqrt{\frac{\rho_{\mathrm{jet}}}{\rho_{\mathrm{HI}}}}
\end{equation}
\begin{equation}
v_{\mathrm{ws, ISM}} = v_{\mathrm{jet}}\frac{r_{\mathrm{jet}}}{r_{\mathrm{ws}}}\sqrt{\frac{\rho_{\mathrm{jet}}}{\rho_{\mathrm{ISM}}}}
\end{equation}
where $v_{\mathrm{ws, HI}}$ and $v_{\mathrm{ws, ISM}}$ are the propagation speed of the jet in the HI cloud and in the warm ISM, respectively. 
Substituting our simulation parameters in equations 
(9) and (10), we obtain $v_{\mathrm{ws, HI}} = 2.5\ \mathrm{km\ s^{-1}}$ and $v_{\mathrm{ws, ISM}} = 17\ \mathrm{km\ s^{-1}}$ when $r_{\mathrm{ws}}$ is $2\ \mathrm{pc}$. 
The propagation time scale is estimated to be 
\begin{equation}
t = \frac{fL}{v_{\mathrm{ws, HI}}} + \frac{(1-f)L}{v_{\mathrm{ws, ISM}}}
\end{equation}
where $f$ is the volume filling factor of the HI clouds, and $L$ is the length scale of the jet. 
In the region $z<10\ \mathrm{pc}$, since the filling factor of the HI clouds is 0, the propagation time scale is $10\ \mathrm{pc}/v_{\mathrm{ws,ISM}} = 0.56\ \mathrm{Myr}$. 
If we assume that $L$ is the length of the beam at $z>10\ \mathrm{pc}$, $L=38\ \mathrm{pc}$ for model F02, $L=25\ \mathrm{pc}$ for model F08, and $L=35\ \mathrm{pc}$ for model F09. 
We obtain the propagation time scale $t=5.1\ \mathrm{Myr}$ for model F02, $t=8.5\ \mathrm{Myr}$ for model F08, and $t=12.7\ \mathrm{Myr}$ for model F09. 
The mean propagation velocity of the jet, evaluated as  
\begin{equation}
v_{\mathrm{prop}}=L/t = \left( \frac{f}{v_{\mathrm{ws, HI}}} + \frac{1-f}{v_{\mathrm{ws, ISM}}} \right)^{-1}
\end{equation}
for models F02, \tr{F02H, F02L,} F08, and F09 is estimated to be $v_{\mathrm{F02}}=7.9\ \mathrm{km\ s^{-1}}$, \tr{$v_{\mathrm{F02H}}=10.5\ \mathrm{km\ s^{-1}}$, $v_{\mathrm{F02L}}=5.3\ \mathrm{km\ s^{-1}}$,} $v_{\mathrm{F08}}=3.0\ \mathrm{km\ s^{-1}}$, and $v_{\mathrm{F09}}=2.7\ \mathrm{km\ s^{-1}}$, respectively.

\begin{figure}[!th]
    \plotone{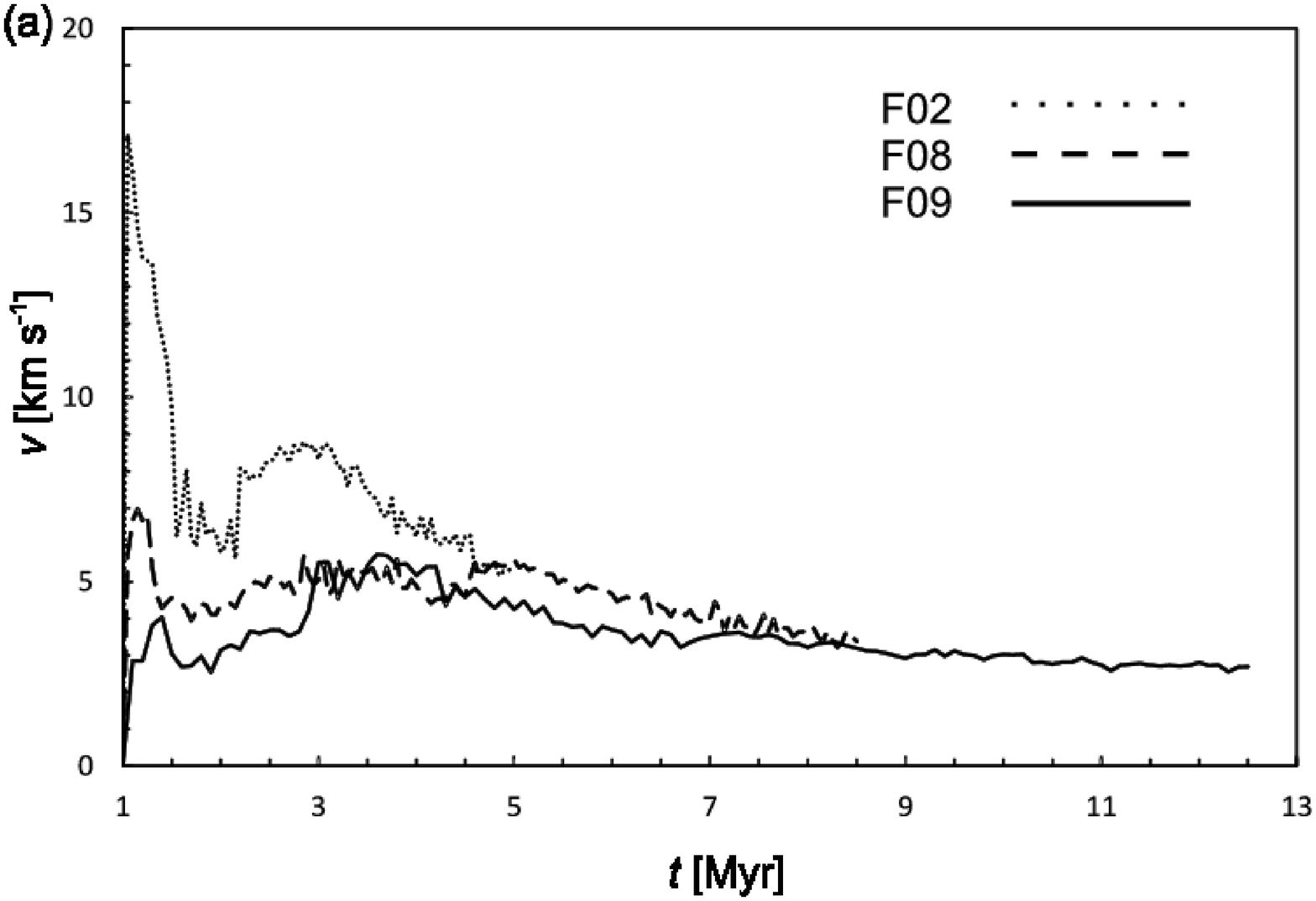}
    \plotone{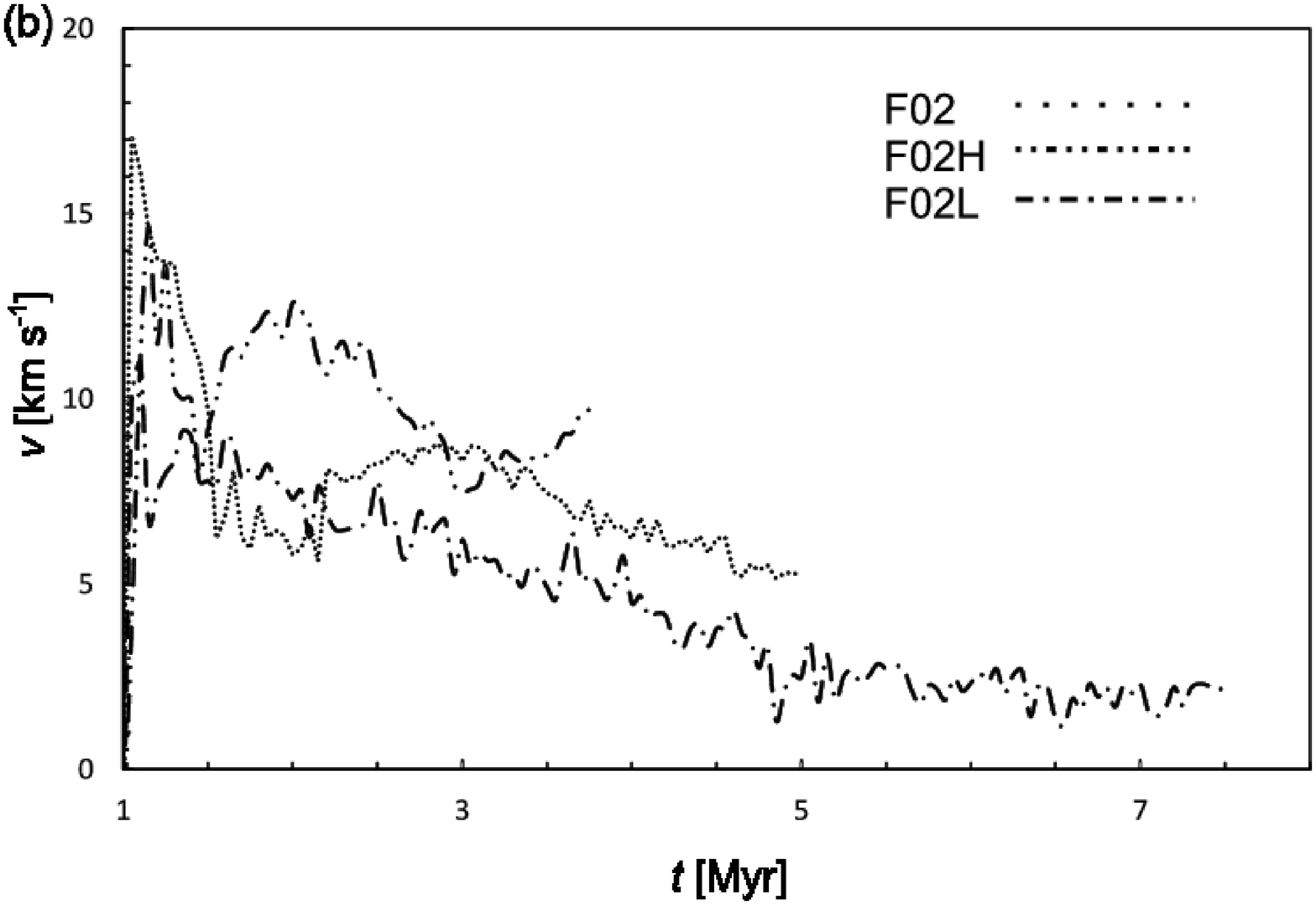}
    \caption{Propagation speed of the jet defined as $(z(t)-z(1\ \mathrm{Myr}))/(t-1\ \mathrm{Myr})$ \tr{for models with \tra{the} same jet speed and (b) models with different jet speed}. The propagation speed approaches to the mean propagation velocity of the jet estimated from equation (12). For model F02 and model F02L, the propagation speed becomes smaller than the estimated value. This is because the jet propagates obliquely after the collision with HI clumps and decrease of the velocity in the head of the jet decrease the ram pressure in the working surface. }
    \label{t-mvp}
\end{figure}

Figure \ref{t-mvp} \tr{(a) and (b)} show the mean propagation speed of the jet after it collides with the HI clumps. 
The mean propagation speed, which is defined as $(z(t)-z(1\ \mathrm{Myr}))/(t-1\ \mathrm{Myr})$, decreases and approaches the propagation speed estimated above. 
In the last stage, it is about $6\ \mathrm{km\ s^{-1}}$ for model F02, \tr{$10\ \mathrm{km\ s^{-1}}$ for model F02H, $2\ \mathrm{km\ s^{-1}}$ for model F02L,} and $3\ \mathrm{km\ s^{-1}}$ for models F08 and F09. 
For \tr{model F02H,} model F08 and model F09, the propagation velocity is comparable to the value estimated by equation (12). 
For model F02 \tr{and model F02L}, \tr{they} become smaller than the estimated values of $7.9\ \mathrm{km\ s^{-1}}$ \tr{and $5.3\ \mathrm{km\ s^{-1}}$}. 
This is because the jet propagates obliquely after colliding with the HI clumps at $z=30\ \mathrm{pc}$, and the decrease in the velocity in the head of the jet decreases the ram pressure at the working surface. 

\begin{figure}[!ht]
  \epsscale{0.7}
   \plotone{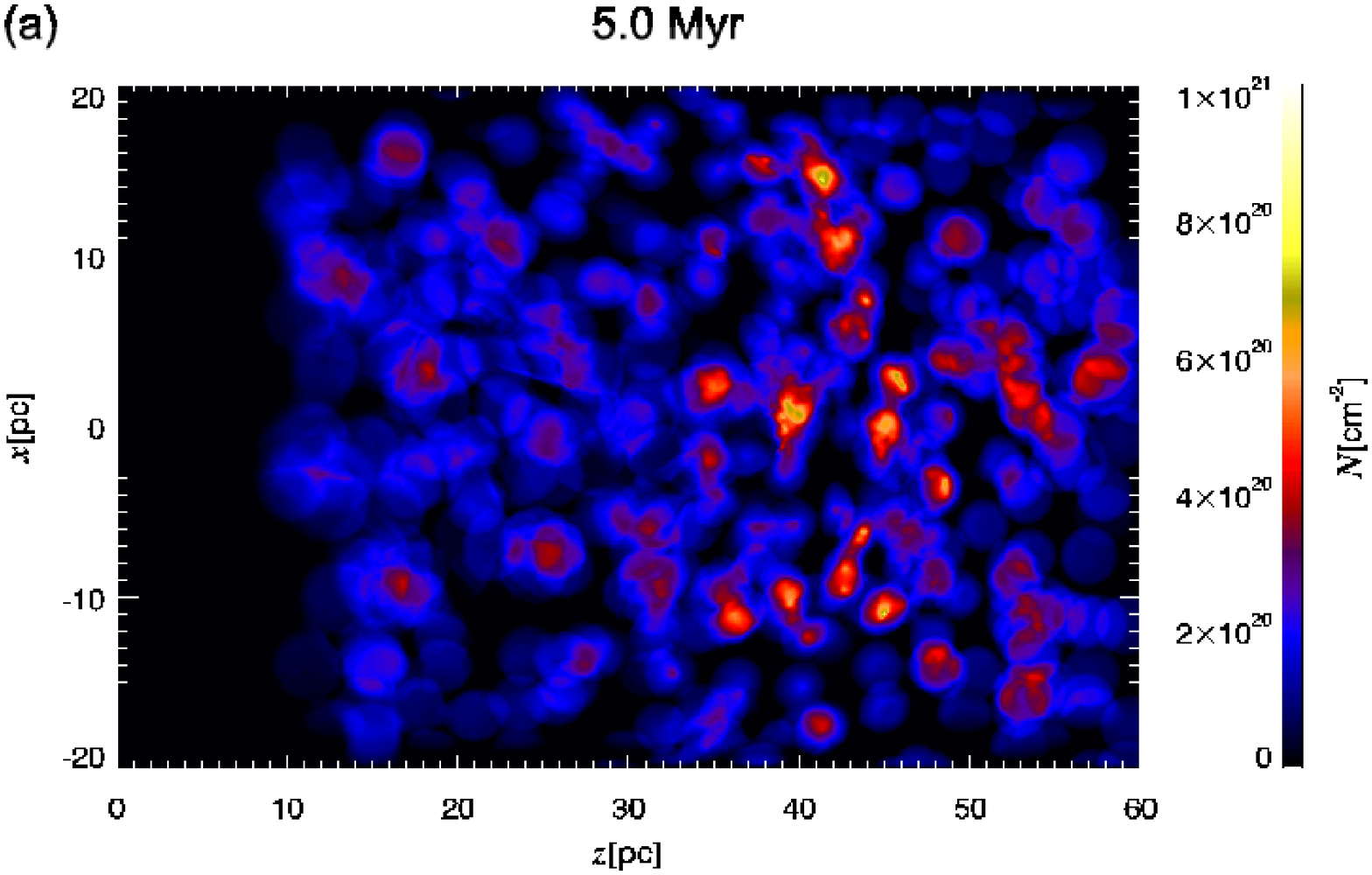}
   \plotone{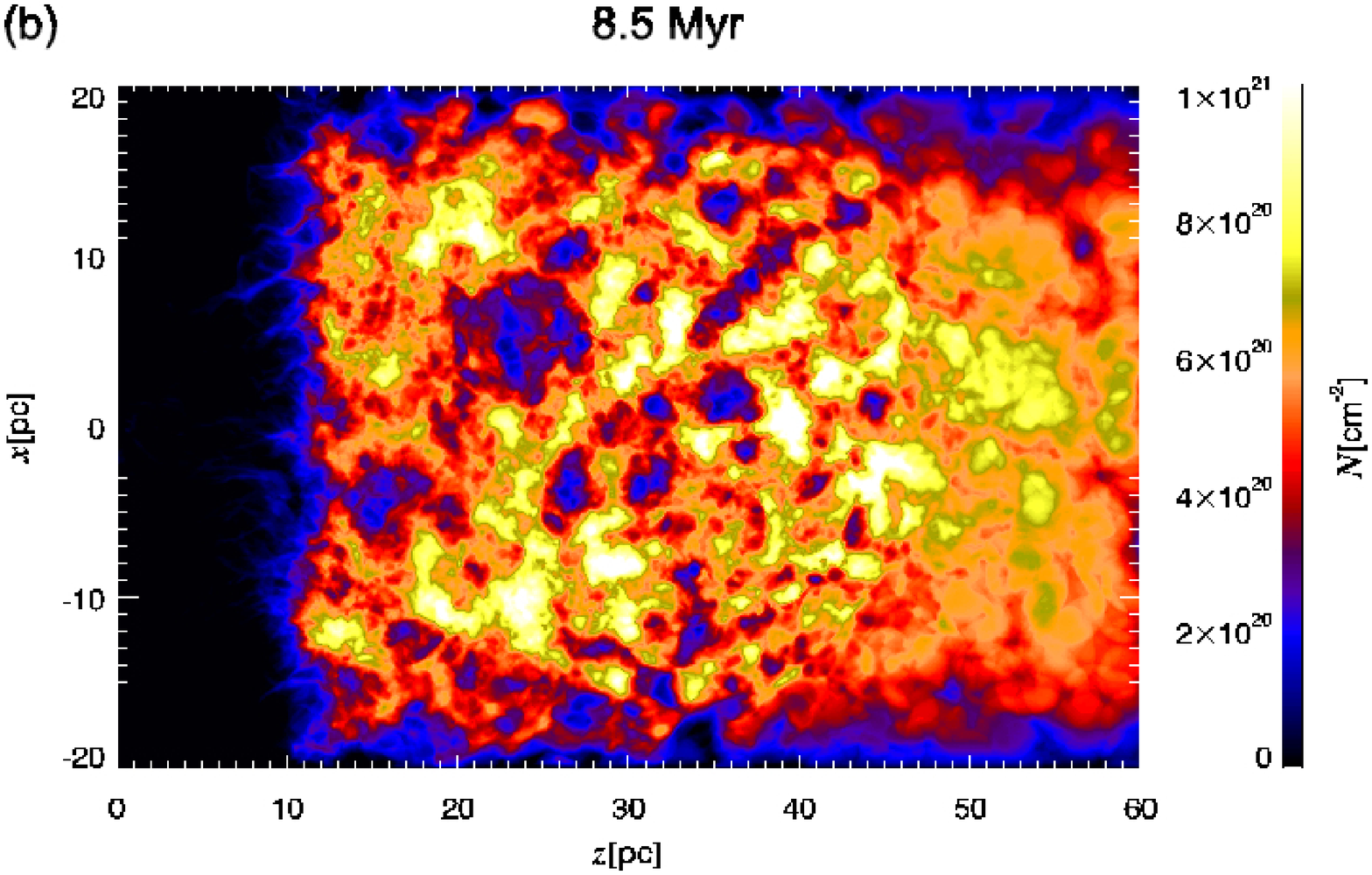}
   \plotone{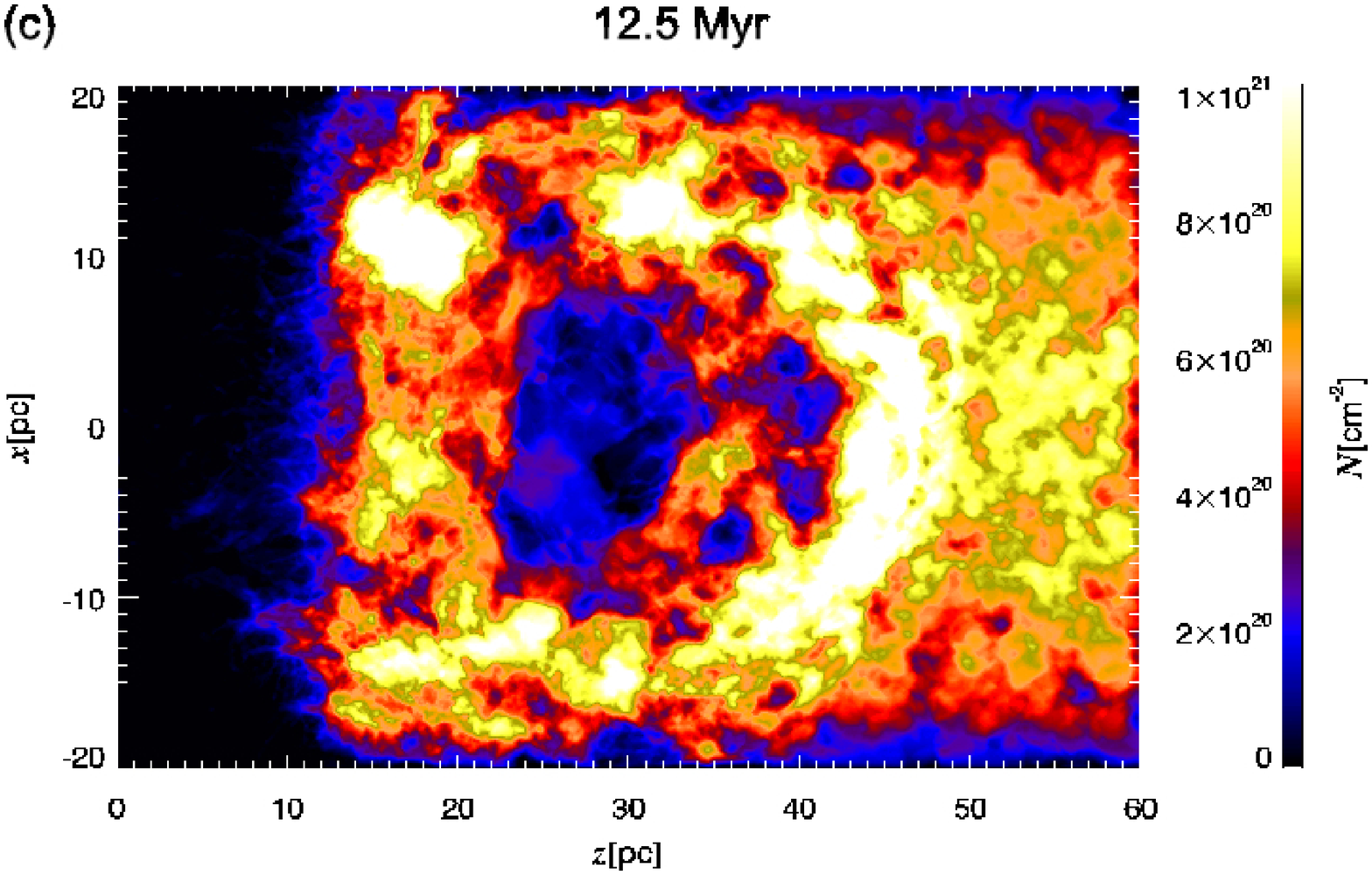}
  \caption{Column number density observed from the azimuth angle $\phi = 140^{\circ}$ for (a) model F02, (b) model F08, and (c) model F09. The distribution becomes broader than the big HI cloud model. For model F09, HI-cavity is formed by the jet sweeping HI clumps. The peak column number density is about $10^{21}\ \mathrm{cm^{-2}}$. \label{wl2f10}}
\end{figure}

Figure \ref{wl2f10} (a), (b), and (c) show the column number density observed from the azimuth angle $\phi = 140^{\circ}$ for models F02, F08, and F09, respectively. 
Shock compression by the jet occurs in various places where the jet sweeps the clouds. 
Therefore, cold, dense clouds distribute more widely than in the large HI cloud model (see Figure \ref{wl2f6} \trb{and Figure \ref{lc2sig}}). 
For model F09, an HI cavity is formed in the region $10\ \mathrm{pc} < z < 30\ \mathrm{pc}$. 
The peak column number density is on the order of $10^{21}\ \mathrm{cm^{-2}}$, which is consistent with that in HI observations. 

\begin{figure}[!ht]
  \epsscale{0.7}
   \plotone{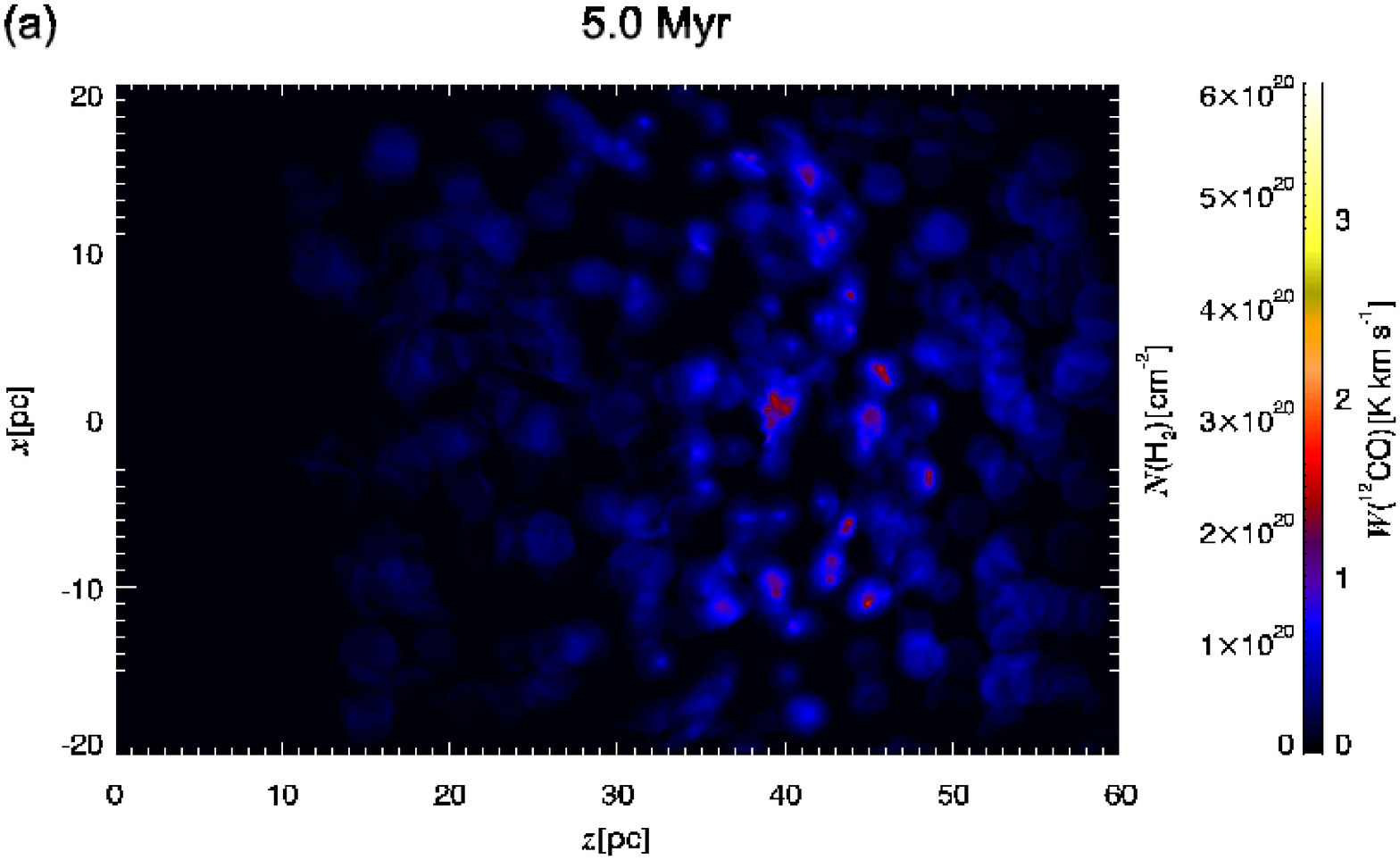}
   \plotone{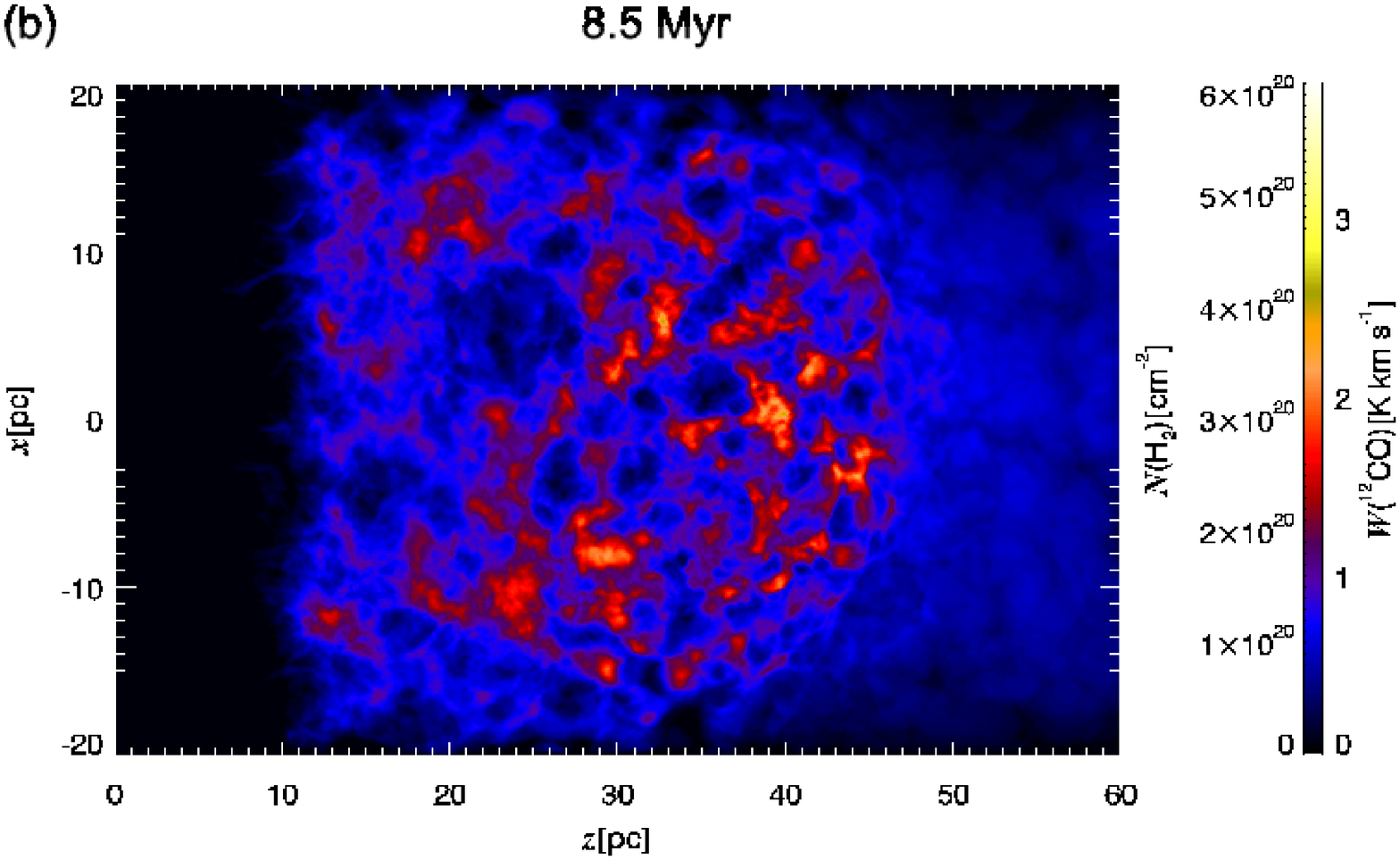}
   \plotone{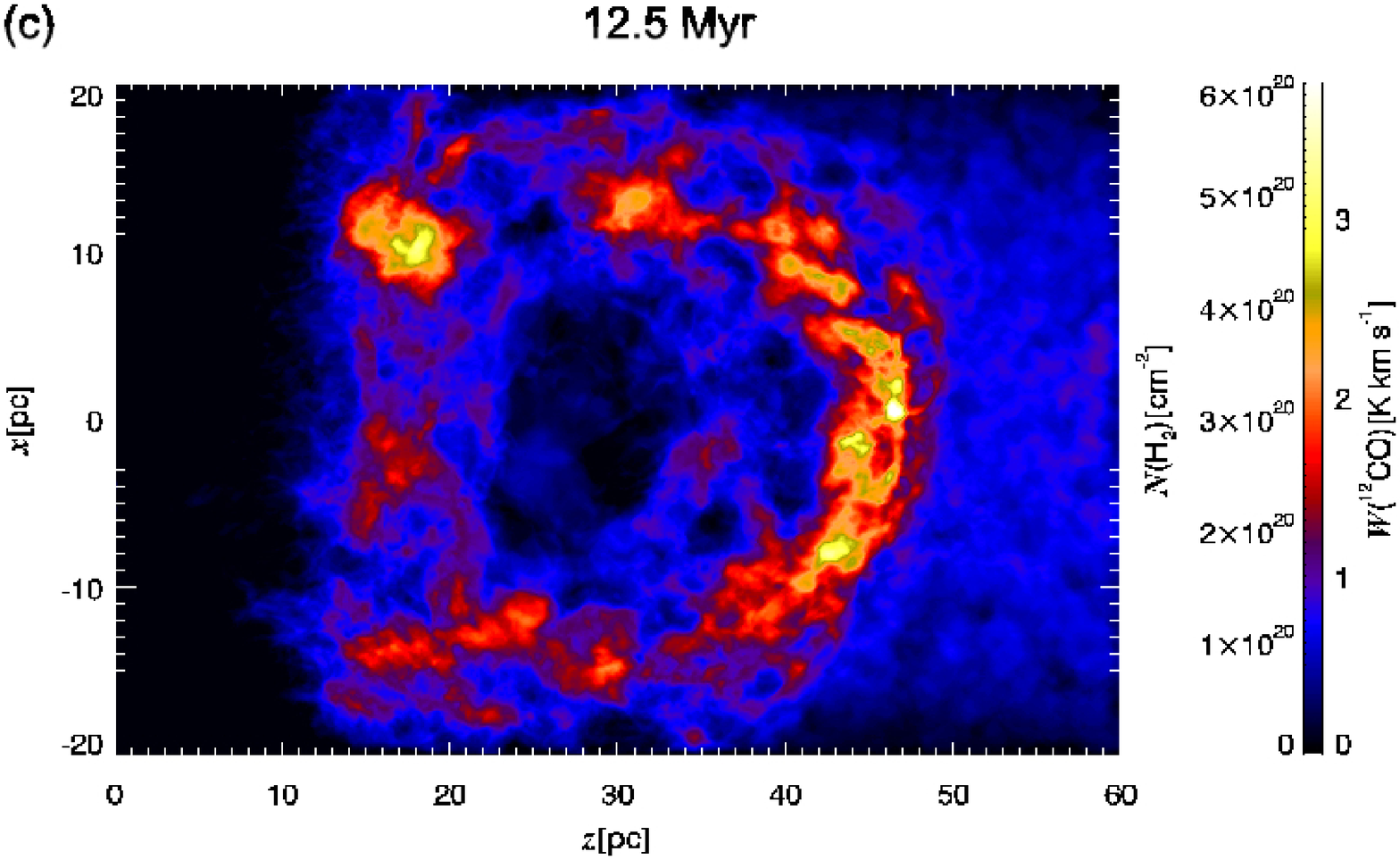}
  \caption{Column number density of $\mathrm{H_{2}}$ and $^{12}$CO$(J=1-0)$ intensity observed from the azimuth angle $\phi = 140^{\circ}$ for (a) model F02, (b) model F08, and (c) model F09. The peak of the $\mathrm{H_{2}}$ column number density is formed in the region where the jet collides with the HI clumps and is about $0.4 \times 10^{21}\ \mathrm{cm^{-2}}$. \label{wl2f11}}
\end{figure}

Figure \ref{wl2f11} (a), (b), and (c) show the column number density of $\mathrm{H_{2}}$ and the CO intensity observed from the azimuth angle $\phi = 140^{\circ}$.
The peak of the CO intensity is located in the regions where the jet swept the HI clouds at $30\ \mathrm{pc}<z<50\ \mathrm{pc}$ for model F02, $30\ \mathrm{pc}<z<35\ \mathrm{pc}$ for model F08, and $35\ \mathrm{pc}<z<45\ \mathrm{pc}$ for model F09. 
The peak CO intensity is about $2.5\ \mathrm{K\ km\ s^{-1}}$, which is smaller than that in CO observations. 
For model F09, the CO distribution approaches that in the arc clouds. 
The numerical results indicate that as the filling factor increases, the distribution of the molecular clouds approaches that in the arc-like clouds. 
The dispersion of the line-of-sight velocity is $1-2\ \mathrm{km\ s^{-1}}$, which is comparable to that in the large HI cloud model and is half that in observations of the jet cloud. 

\trc{
  Let us compare the numerical results for clumpy HI clouds model F09 shown in Figure 13 with the off-center collision model shown in Figure 6.
  The arc-like shape of the dense, cold region, peak column number density, and the slightly asymmetric distribution of the column number density in model F09 are similar to those in the off-center collision model but the dense region is more clumpy and wider in the clumpy clouds model than in the off-center collision with a large cloud. 
}

We would like to estimate the filling factor of HI clouds on the side of the jet clouds by applying equation (11). 
The jet propagation time scale for the jet cloud should be the same as that for the arc cloud, and is given by 
\begin{equation}
\left( \frac{f_{\mathrm{jc}}}{v_{\mathrm{ws,HI}}} + \frac{1-f_{\mathrm{jc}}}{v_{\mathrm{ws,ISM}}} \right) L_{\mathrm{jc}} = \left( \frac{f_{\mathrm{ac}}}{v_{\mathrm{ws,HI}}} + \frac{1-f_{\mathrm{ac}}}{v_{\mathrm{ws,ISM}}} \right) L_{\mathrm{ac}}
\end{equation}
where subscripts ``jc'' and ``ac'' denote the parameters for the jet cloud and arc cloud, respectively. 
We assumed that the jet parameters and density of the ISM and HI cloud are the same for the jet cloud and arc cloud. 
If we solve for $f_{\mathrm{jc}}$ and use equation (9) and (10), equation (13) becomes 
\begin{eqnarray}
f_{\mathrm{jc}} & = & f_{\mathrm{ac}} \frac{L_{\mathrm{ac}}}{L_{\mathrm{jc}}} + \left(1-\frac{L_{\mathrm{ac}}}{L_{\mathrm{jc}}}\right) \left(\frac{v_{\mathrm{ws,ISM}}}{v_{\mathrm{ws,HI}}}-1 \right)^{-1} \\
& = & f_{\mathrm{ac}} \frac{L_{\mathrm{ac}}}{L_{\mathrm{jc}}} + \left(1-\frac{L_{\mathrm{ac}}}{L_{\mathrm{jc}}}\right) \left(\sqrt{\frac{\rho_{\mathrm{HI}}}{\rho_{\mathrm{ISM}}}}-1\right)^{-1}
\end{eqnarray}
This equation does not depend on the jet parameters. 
When $L_{\mathrm{jc}}=90\ \mathrm{pc}$, $L_{\mathrm{ac}}=40\ \mathrm{pc}$, assuming that the density of the ISM and HI clouds are the same as that in our simulations, we obtain
\begin{equation}
f_{\mathrm{jc}} \sim 0.44 f_{\mathrm{ac}} + 0.17
\end{equation}
Since $0 \leq f_{\mathrm{ac}} \leq 1$, the range of $f_{\mathrm{jc}}$ is $0.17 \leq f_{\mathrm{jc}} \leq 0.61$. 
The numerical results indicate that when $f_{\mathrm{ac}} > 0.9$, an arc-like molecular cloud is formed. 
As a result, we obtain $0.57 \leq f_{\mathrm{jc}} \leq 0.61$. 
The ages of the jet cloud and arc cloud are estimated to be a few million years. 
Substituting $t = 2\ \mathrm{Myr}, r_{\mathrm{ws}}=2\ \mathrm{pc}, f=0.6, L=90\ \mathrm{pc}$, and the density of the jet assumed in our simulation in equation\tr{s(9), (10), and} (11), we obtain $v_{\mathrm{jet}} \sim 6.6 \times 10^{3}\ \mathrm{km\ s^{-1}}$.
The kinetic energy injected by the jet is of the order of $10^{36}\ \mathrm{erg\ s^{-1}}$. 

\section{Summary and Discussion}
We carried out 3D MHD simulations of the interaction of the jet with the interstellar HI clouds for two models; a large HI cloud and clumpy HI clouds. 
The density distribution of the HI clouds affects jet propagation and the shape of the cold, dense clouds formed by the jet-cloud interaction. 
When the jet collides with the large HI cloud, the jet sweeps the HI gas and the arc-like cold, dense cloud and HI-cavity are formed. 
When the jet interacts with the clumpy HI clouds, the jet breaks up into branches and the cold, dense clouds distribute more broadly than the large HI cloud model. 
When the volume filling factor is large, the HI-cavity is formed and the shape of the dense clouds approaches the arc-like cloud. 
The density distribution of the interstellar HI clouds determines the shape of the molecular clouds formed by the jet compression. 

Let us compare our results with CO observations \citep{2014ApJ...781...70F}.
The velocity dispersion of the molecular clouds formed by the jet-cloud interaction depends on the jet speed. 
From equation (9), the radial velocity is proportional to $v_{\mathrm{jet}}\sqrt{\rho_{\mathrm{jet}}/\rho_{\mathrm{HI}}}=\sqrt{2E_{\mathrm{jet}}/\rho_{\mathrm{HI}}}$, where $E_{\mathrm{jet}}$ is the kinetic energy density of the jet. 
In our simulation, $E_{\mathrm{jet}}=1.42\times10^{-12}\ \mathrm{erg\ cm^{-3}}, \rho_{\mathrm{HI}}=1.15 \times 10^{-23}\ \mathrm{g\ cm^{-3}}$, the jet speed $v_{\mathrm{jet}}=5.8\times 10^{2}\ \mathrm{km\ s^{-1}}$ and the kinetic energy flux is $2.34 \times 10^{33}\ \mathrm{erg\ s^{-1}}$. 
The velocity dispersion of molecular clouds obtained by numerical simulations when $E_{\mathrm{jet}}/\rho_{\mathrm{HI}}=1.23 \times 10^{11}\ \mathrm{erg\ g^{-1}}$ is about $2\ \mathrm{km\ s^{-1}}$. 
To explain the \tra{observed} velocity dispersion of $\sim 4\ \mathrm{km\ s^{-1}}$, we need $E_{\mathrm{jet}}/\rho_{\mathrm{HI}} \sim 5.0 \times 10^{11}\ \mathrm{erg\ g^{-1}}$ if the velocity dispersion is produced by the jet. 
For this energy and density, the jet speed is estimated to be about $6.6\times 10^{3}\ \mathrm{km\ s^{-1}}$ and the kinetic energy flux of the jet is the order of $10^{36}\ \mathrm{erg\ s^{-1}}$ in our model. 
Another origin of the velocity dispersion of the molecular clouds is the velocity dispersion of the HI clumps. 
The shocked HI clumps can form the molecular clouds with the velocity dispersion which is comparable to the velocity dispersion of the HI clumps when the effect of the jet is small for the velocity of the molecular clouds. 
The kinetic energy of the jet is limited to $E_{\mathrm{jet}}/\rho_{\mathrm{HI}} < 4.9 \times 10^{11}\ \mathrm{erg\ g^{-1}}$ since the velocity dispersion of the molecular clouds produced by the jet need to be smaller than the velocity dispersion of the HI clumps.

\tr{
Let us discuss the dependence on the jet velocity.
In this paper, we assumed that $v_{\mathrm{jet}} = 5.8\times 10^{2}\ \mathrm{km\ s^{-1}}$ \tr{for fiducial models}.
This velocity corresponds to Mach 50 for the warm ISM and Mach 350 for HI clouds. 
\tr{The dependence on the jet speed is studied by carrying out} \tra{simulations} for Mach 4 jet (model F02H) \tr{and Mach 2 jet (model F02L)} in section 3. 
Let us discuss the dependence on the jet speed faster than Mach 4.
Around the jet head, the velocity dispersion of the cold clumps is proportional to $v_{\mathrm{jet}}$ (see equation (9)). 
The cold clumps are no longer accelerated by the ram pressure of the jet after the jet head crosses the clump.
In this stage, we can estimate the velocity dispersion by using equation (18) in \cite{2014ApJ...789...79A}.
The velocity of cold clumps is roughly proportional to $v_{\mathrm{jet}}^{1/3}$.
For instance, the velocity dispersion can be about $200\ \mathrm{km\ s^{-1}}$ around the jet head
if the jet velocity is $0.2c$ which is about 100 times larger than that in our simulations.
Outside the jet head, the velocity dispersion can be $4-8\ \mathrm{km\ s^{-1}}$. 
This velocity dispersion is consistent with observations. 
\tr{\cite{2014ApJ...789...79A} showed that the shape of the interface between the cocoon and the cold sheath is independent of the jet velocity if the jet length is the same. 
Therefore, the shape of the cold clouds can be almost the same even if we change the jet velocity when the filling factor is large. 
When the filling factor is moderate or small, the shape of the cold clouds depends on how much the jet is \tr{reflected} by the HI clumps. 
When the jet velocity is large, the cold clouds can distribute linearly around the jet axis since it is hard for the jet to be deflected by collision with clumps. 
}}

The radius of the HI clumps does not affect the propagation of the jet when the filling factor is close to $0$ or $1$. 
However, it can affect the propagation of the jet when the filling factor is moderate. 
When the \tr{width} of the HI clumps is smaller than the radius of the working surface of the jet, the jet propagates in channels between the HI clumps, and breaks up into numerous branches. 
When the radius of the HI clumps is larger than the radius of the working surface, the jet approaches to that for the large filling factor models. 

\tr{
In this paper, we assumed that the HI clumps are not threaded by the magnetic field at the initial state. 
When the magnetic field is strong, shocked HI clumps are hard to shrink 
in the direction perpendicular to the magnetic field. 
However\tra{,} they can shrink along the magnetic field. 
The preexisting magnetic field can affect the shape of the cold clumps. 
}

HESS J1023-575 is observed both at the high energy band (above $2.5$ TeV) and the low energy band ($0.7-2.5$ TeV). 
The cooling by the pp-interaction and diffusion time scales of the relativistic particles are estimated to be about $5\times 10^{6}\ \mathrm{yr}$ and $2\times 10^{4}\ \mathrm{yr}$, respectively, by \cite{2014ApJ...781...70F}. 
If the arc and the jet clouds and TeV $\gamma$-ray source are produced by a single object, a supernova remnant or a pulsar wind nebulae is not likely as the origin for TeV $\gamma$-ray source since the diffusion time scale of the relativistic particles is shorter than the age of the arc and jet clouds ($\sim \mathrm{Myr}$). 
A microquasar jet which is active over $1\ \mathrm{Myr}$ can explain both the age of the arc and jet clouds and the lifetime of the relativistic particles. 
They discussed that averaged power injected to the relativistic protons need to be $\sim 10^{37}\ \mathrm{erg\ s^{-1}}$ when the diffusion time is $2\times 10^{4}\ \mathrm{yr}$ in order to explain the energy of TeV $\gamma$-ray emission. 
At the low energy band ($0.7-2.5$ TeV), TeV $\gamma$-ray is also detected toward the jet clouds. 
The cooling and diffusion time scales are estimated to be $6\times 10^{4}\ \mathrm{yr}$ and $7\times 10^{3}\ \mathrm{yr}$, respectively. 
Assuming the diffusion time scale of $7\times 10^{3}\ \mathrm{yr}$, the averaged power injection need to be $\sim 8\times 10^{33}\ \mathrm{erg\ s^{-1}}$. 
These estimations indicate that we need to consider the higher energy jet than that in our simulations. 

The peak of the column number density is about $10^{21}\ \mathrm{cm^{-2}}$ in our simulations while the column number density estimated from HI $21\ \mathrm{cm}$ line observations \citep{2005ApJS..158..178M} is about $1.8 \times 10^{21}\ \mathrm{cm^{-2}}$. 
The peaks of the $\mathrm{H_{2}}$ column number density of the arc cloud and jet cloud are $2.7 \times 10^{21}\ \mathrm{cm^{-2}}$ and $1.7-4.8 \times 10^{21}\ \mathrm{cm^{-2}}$, respectively in observations. 
It is about $0.4 \times 10^{21}\ \mathrm{cm^{-2}}$ in our simulations for both the big HI cloud model and clumpy HI clouds model. 
One reason why numerical results gave smaller $\mathrm{H}_{2}$ column density is that the density of the HI clouds is smaller in our simulations. 
For example, when we consider that the HI clouds is twice as dense as the HI clouds in simulations, the peak of the column number density can become twice and the peak of the $\mathrm{H_{2}}$ column number density becomes about four times since we estimate from \cite{2014MNRAS.440.3349R} that the ratio of $\mathrm{H_{2}}$ with the total number density is roughly proportional to the total number density in $1\ \mathrm{cm^{-2}}<n<10^{2}\ \mathrm{cm^{-2}}$. 
Another reason can be uncertainty of the X-factor. 
In this paper, we used $N(\mathrm{H_{2}})/W(^{12}\mathrm{CO})=1.6\times 10^{20}\ \mathrm{cm^{-2}/(K\ km\ s^{-1})}$ \citep{1997ApJ...481..205H} 
obtained by the Energetic Gamma-Ray Experiment Telescope (EGRET) observations. 
X-factor is estimated to be $2.8\times 10^{20}\ \mathrm{cm^{-2}/(K\ km\ s^{-1})}$ from $\gamma$-ray observations \citep{1986A&A...154...25B}, $(1.9\pm1.1)\times 10^{20}\ \mathrm{cm^{-2}/(K\ km\ s^{-1})}$ from infrared observations \citep{1998ApJ...507..507R}, and so on. 
Thus X-factor has uncertainty in the range $1-3 \times 10^{20}\ \mathrm{cm^{-2}/(K\ km\ s^{-1})}$. 

We thank T. Hanawa, Y. Matsumoto, S. Miyaji, J. M. Stone, M. Machida, A. Mizuta and T. Minoshima for discussion. 
Numerical computations were carried out by XC30 at Center for Computational Astrophysics, NAOJ. 
This work is supported by grants in aid for scientific research by JSPS KAKENHI Grant Number (21253003, 23340040, 16H03954) and Grant-in-Aid for JSPS Fellow (24.4786).

\bigskip

\bibliography{apj-jour,letterapj}

\begin{thebibliography}{26}
\expandafter\ifx\csname natexlab\endcsname\relax\def\natexlab#1{#1}\fi

\bibitem[{{Asahina} {et~al.}(2014){Asahina}, {Ogawa}, {Kawashima}, {Furukawa},
  {Enokiya}, {Yamamoto}, {Fukui}, \& {Matsumoto}}]{2014ApJ...789...79A}
{Asahina}, Y., {Ogawa}, T., {Kawashima}, T., {et~al.} 2014, \apj, 789, 79

\bibitem[{{Begelman} \& {McKee}(1990)}]{1990ApJ...358..375B}
{Begelman}, M.~C., \& {McKee}, C.~F. 1990, \apj, 358, 375

\bibitem[{{Bloemen} {et~al.}(1986){Bloemen}, {Strong}, {Mayer-Hasselwander},
  {Blitz}, {Cohen}, {Dame}, {Grabelsky}, {Thaddeus}, {Hermsen}, \&
  {Lebrun}}]{1986A&A...154...25B}
{Bloemen}, J.~B.~G.~M., {Strong}, A.~W., {Mayer-Hasselwander}, H.~A., {et~al.}
  1986, \aap, 154, 25

\bibitem[{{Blondin} {et~al.}(1990){Blondin}, {Fryxell}, \&
  {K{\"o}nigl}}]{1990ApJ...360..370B}
{Blondin}, J.~M., {Fryxell}, B.~A., \& {K{\"o}nigl}, A. 1990, \apj, 360, 370

\bibitem[{{Blondin} {et~al.}(1989){Blondin}, {K{\"o}nigl}, \&
  {Fryxell}}]{1989ApJ...337L..37B}
{Blondin}, J.~M., {K{\"o}nigl}, A., \& {Fryxell}, B.~A. 1989, \apjl, 337, L37

\bibitem[{{Borges} {et~al.}(2008){Borges}, {Carmona}, {Costa}, \&
  {Don}}]{2008JCoPh.227.3191B}
{Borges}, R., {Carmona}, M., {Costa}, B., \& {Don}, W.~S. 2008, Journal of
  Computational Physics, 227, 3191

\bibitem[{{Clarke} {et~al.}(1986){Clarke}, {Norman}, \&
  {Burns}}]{1986ApJ...311L..63C}
{Clarke}, D.~A., {Norman}, M.~L., \& {Burns}, J.~O. 1986, \apjl, 311, L63

\bibitem[{{Frank} {et~al.}(1998){Frank}, {Ryu}, {Jones}, \&
  {Noriega-Crespo}}]{1998ApJ...494L..79F}
{Frank}, A., {Ryu}, D., {Jones}, T.~W., \& {Noriega-Crespo}, A. 1998, \apjl,
  494, L79

\bibitem[{{Fukui} {et~al.}(2009){Fukui}, {Furukawa}, {Dame}, {Dawson},
  {Yamamoto}, {Rowell}, {Aharonian}, {Hofmann}, {de O{\~n}a Wilhelmi},
  {Minamidani}, {Kawamura}, {Mizuno}, {Onishi}, {Mizuno}, \&
  {Nagataki}}]{2009PASJ...61L..23F}
{Fukui}, Y., {Furukawa}, N., {Dame}, T.~M., {et~al.} 2009, \pasj, 61, L23

\bibitem[{{Furukawa} {et~al.}(2014){Furukawa}, {Ohama}, {Fukuda}, {Torii},
  {Hayakawa}, {Sano}, {Okuda}, {Yamamoto}, {Moribe}, {Mizuno}, {Maezawa},
  {Onishi}, {Kawamura}, {Mizuno}, {Dawson}, {Dame}, {Yonekura}, {Aharonian},
  {de O{\~n}a Wilhelmi}, {Rowell}, {Matsumoto}, {Asahina}, \&
  {Fukui}}]{2014ApJ...781...70F}
{Furukawa}, N., {Ohama}, A., {Fukuda}, T., {et~al.} 2014, \apj, 781, 70

\bibitem[{{Hunter} {et~al.}(1997){Hunter}, {Bertsch}, {Catelli}, {Dame},
  {Digel}, {Dingus}, {Esposito}, {Fichtel}, {Hartman}, {Kanbach}, {Kniffen},
  {Lin}, {Mayer-Hasselwander}, {Michelson}, {von Montigny}, {Mukherjee},
  {Nolan}, {Schneid}, {Sreekumar}, {Thaddeus}, \&
  {Thompson}}]{1997ApJ...481..205H}
{Hunter}, S.~D., {Bertsch}, D.~L., {Catelli}, J.~R., {et~al.} 1997, \apj, 481,
  205

\bibitem[{{Inoue} {et~al.}(2006){Inoue}, {Inutsuka}, \&
  {Koyama}}]{2006ApJ...652.1331I}
{Inoue}, T., {Inutsuka}, S.-i., \& {Koyama}, H. 2006, \apj, 652, 1331

\bibitem[{{K{\"o}ssl} {et~al.}(1990){K{\"o}ssl}, {M{\"u}ller}, \&
  {Hillebrandt}}]{1990A&A...229..378K}
{K{\"o}ssl}, D., {M{\"u}ller}, E., \& {Hillebrandt}, W. 1990, \aap, 229, 378

\bibitem[{{Lind} {et~al.}(1989){Lind}, {Payne}, {Meier}, \&
  {Blandford}}]{1989ApJ...344...89L}
{Lind}, K.~R., {Payne}, D.~G., {Meier}, D.~L., \& {Blandford}, R.~D. 1989,
  \apj, 344, 89

\bibitem[{{McClure-Griffiths} {et~al.}(2005){McClure-Griffiths}, {Dickey},
  {Gaensler}, {Green}, {Haverkorn}, \& {Strasser}}]{2005ApJS..158..178M}
{McClure-Griffiths}, N.~M., {Dickey}, J.~M., {Gaensler}, B.~M., {et~al.} 2005,
  \apjs, 158, 178

\bibitem[{{Minoshima} {et~al.}(2015){Minoshima}, {Hirose}, \&
  {Sano}}]{2015ApJ...808...54M}
{Minoshima}, T., {Hirose}, S., \& {Sano}, T. 2015, \apj, 808, 54

\bibitem[{{Miyoshi} \& {Kusano}(2005)}]{2005JComp...208...315}
{Miyoshi}, T., \& {Kusano}, K. 2005, J. Comp. Phys, 208, 315

\bibitem[{{Norman} {et~al.}(1982){Norman}, {Winkler}, {Smarr}, \&
  {Smith}}]{1982A&A...113...285N}
{Norman}, M.~L., {Winkler}, K.-H.~A., {Smarr}, L., \& {Smith}, M.~D. 1982,
  \aap, 113, 285

\bibitem[{{Reach} {et~al.}(1998){Reach}, {Wall}, \&
  {Odegard}}]{1998ApJ...507..507R}
{Reach}, W.~T., {Wall}, W.~F., \& {Odegard}, N. 1998, \apj, 507, 507

\bibitem[{{Richings} {et~al.}(2014){Richings}, {Schaye}, \&
  {Oppenheimer}}]{2014MNRAS.440.3349R}
{Richings}, A.~J., {Schaye}, J., \& {Oppenheimer}, B.~D. 2014, \mnras, 440,
  3349

\bibitem[{{Stone} \& {Hardee}(2000)}]{2000ApJ...540..192S}
{Stone}, J.~M., \& {Hardee}, P.~E. 2000, \apj, 540, 192

\bibitem[{{Suresh} \& {Huynh}(1997)}]{1997JCoPh.136...83S}
{Suresh}, A., \& {Huynh}, H.~T. 1997, Journal of Computational Physics, 136, 83

\bibitem[{{Te{\c s}ileanu} {et~al.}(2008){Te{\c s}ileanu}, {Mignone}, \&
  {Massaglia}}]{2008A&A...488..429T}
{Te{\c s}ileanu}, O., {Mignone}, A., \& {Massaglia}, S. 2008, \aap, 488, 429

\bibitem[{{Todo} {et~al.}(1992){Todo}, {Uchida}, {Sato}, \&
  {Rosner}}]{1992PASJ...44...245T}
{Todo}, Y., {Uchida}, Y., {Sato}, T., \& {Rosner}, R. 1992, \pasj, 44, 245

\bibitem[{{Todo} {et~al.}(1993){Todo}, {Uchida}, {Sato}, \&
  {Rosner}}]{1993ApJ...403...164}
---. 1993, \apj, 403, 164

\bibitem[{{Yamamoto} {et~al.}(2008){Yamamoto}, {Ito}, {Ishigami}, {Fujishita},
  {Kawase}, {Kawamura}, {Mizuno}, {Onishi}, {Mizuno}, {McClure-Griffiths}, \&
  {Fukui}}]{2008PASJ...60..715Y}
{Yamamoto}, H., {Ito}, S., {Ishigami}, S., {et~al.} 2008, \pasj, 60, 715

\end{thebibliography}

\newpage

\end{document}